\documentclass[a4paper,11pt]{article}
\pdfoutput=1
\usepackage{jheppub} 
\usepackage[T1]{fontenc} 
\usepackage{graphicx}
\usepackage{dcolumn}
\usepackage{bm}
\usepackage{rotating}
\usepackage{color}
\title{\boldmath Adaptive asymptotic solutions of inflationary models in the Hamilton-Jacobi formalism: Application to T-models}
\date{\today}
\author[a,1]{Elena Medina\note{Corresponding author.}}
\author[b]{and Gabriel \'Alvarez}
\affiliation[a]{Departamento de Matem\'aticas,
                        Facultad de Ciencias,
                        Universidad de C\'adiz,
                        11510 Puerto Real, C\'adiz, Spain}
\affiliation[b]{Departamento de F\'{\i}sica Te\'orica,
                        Facultad de Ciencias F\'{\i}sicas,
                        Universidad Complutense,
                        28040 Madrid, Spain}
\emailAdd{elena.medina@uca.es}
\emailAdd{galvarez@ucm.es}
\abstract{We develop a method to compute the slow-roll expansion for the Hubble parameter in inflationary
models in a flat Friedmann-Lema\^itre-Robertson-Walker spacetime that is applicable to a wide class of potentials
including monomial, polynomial, or rational functions of the inflaton, as well as polynomial or rational functions
of the exponential of the inflaton. The method, formulated within the Hamilton-Jacobi formalism, adapts the form
of the slow-roll expansion to the analytic form of the inflationary potential, thus allowing
a consistent order-by-order computation amenable to Pad\'e summation. Using T-models as an
example, we show that Pad\'e summation extends the domain of validity of this adapted slow-roll expansion
to the end of inflation. Likewise, Pad\'e summation extends the domain of validity of kinetic-dominance asymptotic expansions
of the Hubble parameter into the fast-roll regime, where they can be matched to the aforesaid Pad\'e-summed
slow-roll expansions. This matching in turn determines the relation between the expansions for the number $N$ of e-folds and allows us
to compute the total amount of inflation as a function of the initial data or, conversely, to select initial data that correspond
to a fixed total amount of inflation. Using the slow-roll stage expansions, we also derive expansions for the corresponding
spectral index $n_s$ accurate to order $1/N^2$, and tensor-to-scalar ratio $r$ accurate to order $1/N^3$ for these T-models.}
\begin{document}
\maketitle
\flushbottom
\section{Introduction\label{sec:intro}}
The dynamical equations for single-field inflationary models defined by a potential $V(\Phi)$ of an inflaton field $\Phi$
in a spatially flat Friedmann-Lema\^itre-Robertson-Walker spacetime can be written as
\begin{equation}
    \label{eq1}
    \ddot{\Phi}+3H\dot{\Phi}+V'(\Phi) = 0,
\end{equation}
\begin{equation}
    \label{eq2}
    3\mathrm{M}_\mathrm{Pl}^2 H^2 = \frac{1}{2}\dot{\Phi}^2+V(\Phi),
\end{equation}
where a prime denotes the derivative of a function with respect to its argument, dots denote derivatives with respect
to the cosmic time $t$, $H$ is the Hubble parameter defined in terms of the scale factor $a(t)$ as
\begin{equation}
	\label{Ha}
	H=\frac{\dot{a}}{a},
\end{equation}
and $\mathrm{M}_\mathrm{Pl}$ is the reduced Planck mass. During inflation, the inflaton $\Phi$ is a monotonic
function of the cosmic time, and the number of e-folds, defined as
\begin{equation}
	\label{Nt}
	N(t) = \log\frac{a(t_\text{end})}{a(t)},
\end{equation}
can be written equivalently as 
\begin{equation}
	\label{Nvarphi}
	N(\Phi) =  \log\frac{a(\Phi_\text{end})}{a(\Phi)},
\end{equation}
and is often used as a time coordinate~\cite{KA13,AK18,CM19,SD20}, where the subindex ``end'' denotes
magnitudes at the end of the inflationary stage, which corresponds to $N(\Phi_\text{end}) = 0$.

The standard slow-roll (SR) approximation follows from neglecting the kinetic term $\dot{\Phi}^2/2$
in eq.~(\ref{eq2}) and substituting the resulting approximation
\begin{equation}
	\label{eq:SSRAP}
	H^2 \approx \frac{V(\Phi)}{3\mathrm{M}_\mathrm{Pl}^2},
\end{equation}
into the differential equation
\begin{equation}
	\label{eq:dndP}
	\frac{d N}{d \Phi} = \frac{1}{2\mathrm{M}_\mathrm{Pl}^2}\frac{H(\Phi)}{H'(\Phi)},
\end{equation} 
with the said initial condition $N(\Phi_\text{end}) = 0$. 
However, increasingly accurate observational data called for a more accurate expansion
of which the standard SR approximation~(\ref{eq:SSRAP}) would be just the leading term.
Liddle, Parsons and~Barrow~\cite{LID94} developed such an expansion in the form,
\begin{equation}
	\label{eq:SSREX}
	H(\Phi)^2
	=
	 \frac{V(\Phi)}{3\mathrm{M}_\mathrm{Pl}^2}
	 \left(1+\frac{1}{3}\epsilon_{1,V}(\Phi)-\frac{1}{3}\epsilon_{1,V}(\Phi)^2+\frac{2}{9}\epsilon_{1,V}(\Phi )\epsilon_{2,V}(\Phi)+\cdots\right),
\end{equation}
where
\begin{eqnarray}
	\label{eq:SSREX1}
	\epsilon_{1,V}(\Phi) & = & \frac{\mathrm{M}_\mathrm{Pl}^2}{2}\left(\frac{V'(\Phi)}{V(\Phi)}\right)^2,\\
	\label{eq:SSREXn}
	\epsilon_{n,V}(\Phi) & = & \mathrm{M}_\mathrm{Pl}^2
	                                  \left(\frac{V'(\Phi)^{n-2}V^{(n)}(\Phi)}{V(\Phi)^{n-1}}\right)^{1/(n-1)},\quad n=2,3,\dots.
\end{eqnarray}
Equation~(\ref{eq:SSREX}) leads to ensuing expansions for the SR parameters $\epsilon_n$, of which there are several definitions.
We adopt the definitions given in Ref.~\cite{LE02} (which, incidentally, do not agree with the definitions in Ref.~\cite{LID94}),
\begin{equation}
	\label{parSR}
	\epsilon_1(\Phi) = 2\mathrm{M}_\mathrm{Pl}^2\left(\frac{H'(\Phi)}{H(\Phi)}\right)^2,
	\quad \epsilon_{n+1}(\Phi)=\frac{d\ln|\epsilon_n(\Phi)|}{dN},\quad n=1,2,\dots.
\end{equation}
For example,
\begin{equation}
	\epsilon_1(\Phi) = \epsilon_{1,V}(\Phi)-\frac{4}{3}\epsilon_{1,V}(\Phi)^2+\frac{2}{3}\epsilon_{1,V}(\Phi)\epsilon_{2,V}(\Phi)+\cdots,
\end{equation}
and the standard SR approximation (i.e., the leading term of the standard SR expansion) would be to take
$\epsilon_1(\Phi)\approx\epsilon_{1,V}(\Phi)$. This is the usual approximation
used to compute the spectral index $n_s$ and the tensor-to-scalar ratio $r$ for a variety of potentials~\cite{MA14}.

The present paper is motivated by the observation that although the computation of 
the standard SR expansion~(\ref{eq:SSREX}) to high order using eqs.~(\ref{eq:SSREX1}) and~(\ref{eq:SSREXn})
seems straightforward, in practice it is not always so. The order parameter in the standard SR expansion for $H(\Phi)$
as given by eqs.~(\ref{eq:SSREX})--(\ref{eq:SSREXn}) appears to be $V(\Phi)^{-1/2}$, but even for the simplest
potentials like monomial potentials, this standard expansion does not lead to a systematic order-by-order series
that can be directly summed using rational approximants, and for more complicated potentials
with concave, exponentially flat plateaus that seem to give results in good agreement with data
from the Planck satellite~\cite{PL18VI,PL18X} for the ratio of the amplitude of tensor perturbations
to the amplitude of scalar perturbations and with the scalar spectral tilt~\cite{KA13,GA15,KA15},
even carrying out the computation at high order is not straightforward.

Therefore, in this paper we develop a method to compute the SR expansion that is applicable to a wide variety of potentials and which,
in essence, adapts the form of the SR expansion to the analytic form of the potential $V(\Phi)$,
thus allowing a consistent and efficient order-by-order computation. The method is formulated within the
Hamilton-Jacobi formalism of Salopek and Bond~\cite{SA90}, and turns out to be applicable to a wide class of potentials, including monomial, polynomial,
or rational functions of the inflaton, as well as to polynomial or rational functions of the exponential of the inflaton, among which we mention
cosmological $\alpha$-attractors~\cite{KA13,FK13,FF14,KA14,KA15,CA15,AK18} and
the subclass of T-models~\cite{CA15,AK18,SD20,GE21} corresponding to K\"ahler superpotentials $f(Z)=Z^m$.

The idea to adapt the SR expansions is to find a suitable variable $F(\Phi)$ that, if used instead of $V(\Phi)$, leads
to systematic order-by-order computations. Note that if we factor out the leading behavior of $H(\Phi)$ by defining
\begin{equation}
	\label{introchange}
	\mathcal{H}(\Phi) = \sqrt{3} \mathrm{M}_\mathrm{Pl} \frac{H(\Phi)}{\sqrt{V(\Phi)}},
\end{equation}
then $\mathcal{H}(\Phi)$ satisfies~\cite{AM20}~(see Ref.~\cite{HAN14} for an equivalent equation for $\log\mathcal{H}$),
\begin{equation}
	\label{introymas}
	\mathcal{H}'(\Phi) = \frac{1}{ \mathrm{M}_\mathrm{Pl}}
	                                 \sqrt{\frac{3}{2}} \sqrt{\mathcal{H}(\Phi)^2-1}
	                                 - \mathcal{V}(\Phi)  \mathcal{H}(\Phi),
\end{equation}
where
\begin{equation}  
	\label{introiu}
 	\mathcal{V}(\Phi) = \frac{V'(\Phi)}{2 V(\Phi)},
\end{equation}
and whenever this function is of the form
\begin{equation}
	\label{eq:qf}
	\mathcal{V}(\Phi) = \mathcal{Q}(F(\Phi)),
\end{equation}
where $F(\Phi)\to\infty$ as $\Phi\to\infty$, and $\mathcal{Q}$ has an ensuing Taylor expansion in $1/F(\Phi)$,
then eq.~(\ref{introymas}) has a formal solution in inverse powers of $F(\Phi)$. We point out that using these
variables $F(\Phi)$ does not lead to an essentially different SR expansion in the sense that if it were possible
to work with the infinitely many terms of the series, both expansions would ultimately be equivalent.
However, using $F(\Phi)$ (i) leads to a systematic expansion in the sense that only the series for $\mathcal{Q}$
is required, and (ii) the unavoidable truncation to a finite
number of terms is done consistently order-by-order in $1/F(\Phi)$ (i.e., higher-order terms do not
contain contributions from lower-order terms), thereby permitting consistent use of summation methods.
We present the method in the main body of the paper by example using T-models
(for which there is a wealth of first-order SR results for comparison), and defer to Appendix~\ref{ap:a}
the specific choices of $F(\Phi)$ that adapt the method to the other families of potentials mentioned earlier
and the first few terms of the resulting expansions for $H(\Phi)$.

T-models, when written in terms of the canonically normalized field $\Phi$, are described by potentials of the form	
\begin{equation}
    \label{Tmodel}
    V(\Phi) = \Lambda\left(\tanh^2\left(\frac{\Phi}{{\mathrm{M}_\mathrm{Pl}}\sqrt{6\alpha}}\right)\right)^m,
\end{equation}
where $\Lambda$, $\alpha$ and $m$ are positive parameters characterizing the particular model.
In the SR approximation, the end of inflation is defined by setting
\begin{equation}
	\epsilon_1(\Phi_\mathrm{end})\approx\epsilon_{1,V}(\Phi_\mathrm{end}) = 1,
\end{equation}
or, explicitly,
\begin{equation}
	\label{eq:Phiend}
	\sinh\left(\sqrt{\frac{2}{3\alpha}}\frac{\Phi_\text{end}}{\mathrm{M}_\mathrm{Pl}}\right)
	=
	\sqrt{\frac{4m^2}{3\alpha}},
\end{equation}
which leads to (see Ref.~\cite{KA13}, Ref.~\cite{SD20} for the particular case $m=1$, or section~\ref{sec:neSR} below)
\begin{equation}
	\label{standardSRPhi}
	N^\text{{(SR)}}(\Phi) = \frac{3\alpha}{4m}\left(
	\cosh\left(\sqrt{\frac{2}{3\alpha}}\frac{\Phi}{\mathrm{M}_\mathrm{Pl}}\right) -\sqrt{1+\frac{4m^2}{3\alpha}}\right).
\end{equation}
However, computing the total amount of inflation in this approach would still require the determination of $\Phi_\text{in}$,
the value of the inflaton at the beginning of inflation. Note that in computing the total amount of inflation in this
approximation there are two possible sources of error: the first comes from ignoring altogether the kinetic term
in the SR stage (i.e., from the fact that eq.~(\ref{standardSRPhi}) is just the leading term of a full SR asymptotic expansion),
while the second is that $\Phi_\text{in}$ typically lies in the fast roll stage, well beyond the range of applicability 
not only of the first order SR approximation but even of higher-order SR approximations. We will show that
the range of applicability of asymptotic series for the Hubble parameter derived in the kinetic-dominance~(KD) stage~\cite{HAN14,HAN19,HER19},
when appropriately summed as discussed below, extends beyond
the KD stage not only into the fast-roll stage, but into the beginning of the SR stage. Therefore the KD series
can be matched to the appropriately summed SR series, thereby allowing the determination of $\Phi_\text{in}$
as a function of the initial conditions and the comparison of the ensuing total amount of inflation with purely SR results.

Since the determination of $\Phi_\text{in}$ will require us to go into the KD stage,
we mention that dynamical systems theory has been used extensively  to obtain global results in inflationary
cosmology~\cite{BE85} and in particular for  these T-models~\cite{CO98,TA14,PA15,AU15,AU15D,AU17}.
For instance, Alho and Uggla~\cite{AU17} use a Poincar\'e compactification of the $(\Phi, \dot{\Phi})$ phase plane
which resolves singular points at infinity and therefore is specially suited to identify all possible asymptotic
behaviors and all orbits connecting critical points. However, Hamilton-Jacobi methods are better suited
to find and match high-order asymptotic solutions of eqs.~(\ref{eq1}) and~(\ref{eq2}) both in the SR stage and in the KD stage.
In fact, Liddle, Parsons and~Barrow~\cite{LID94} used this kind of methods to find asymptotic expansions in the SR stage,
Handley~{et. al.}~\cite{HAN14,HAN19,HER19} in the KD stage, and Mart\'{\i}nez Alonso~{et. al.}~\cite{AM20,ME20,MM21}
for a variety of potentials in either or both stages. Particularly relevant for the present paper is Ref.~\cite{MM21},
in which Medina and Mart\'{\i}nez Alonso find asymptotic expansions for the SR and KD stages for a generalized Starobinsky model,
\begin{equation}
	V(\Phi) =    \Lambda_1 e^{-\sqrt{6}\lambda\Phi/\mathrm{M}_\mathrm{Pl}}
	                    + \Lambda_2 e^{-\sqrt{6}\mu\Phi/\mathrm{M}_\mathrm{Pl}}
	                    + \Lambda_3,
\end{equation}
where all the parameters are positive, $\lambda>\mu$, and $\Lambda_3$ is chosen so that the minimum of the potential
is zero. After rescaling $\varphi = \sqrt{3/2} \Phi/\mathrm{M}_\mathrm{Pl}$, typical expansions for this potential in the KD stage
(cf.~eq.~(47) in Ref.~\cite{MM21}), are asymptotic series in $e^{-\varphi}$ whose coefficients are polynomials
in $e^{-2\lambda\varphi}$ and $e^{-2\mu\varphi}$. However, in the case of T-models the asymptotic expansions in the KD
stage turn out to be series in $e^{-\varphi}$ whose coefficients are not polynomials, but \emph{series} in $e^{-2\varphi/3\sqrt{\alpha}}$,
whose matching to the series in the SR stage has to be done consistently.

The numerical usage of these expansions to obtain high-accuracy results presents the usual challenges of dealing with
asymptotic series. For instance, although partial sums of the SR formal series give an excellent accuracy at large $\Phi$,
they are not accurate enough in a neighborhood of $\Phi_\text{end}$. A common method to extend the range of
applicability of these (truncated) series is to use rational approximants derived from the power series.
In fact, in studying the high-order $\epsilon_{n,V}(\Phi)$
for the quadratic potential, Liddle, Parsons and~Barrow~\cite{LID94} use [1/1] rational, multivariable Canterbury
approximants to increase the accuracy provided by simple partial sums towards the end of inflation.
As summation method for our asymptotic expansions we use Pad\'e approximants~\cite{BGM},
both to obtain accurate results towards the end of the inflation and to show the existence of a certain interval
on which the appropriately summed SR and KD expansions are both valid, can be matched, and
allow the determination of $\Phi_\mathrm{in}$ mentioned in the previous paragraphs.

The layout of the paper is as follows. In section~\ref{sec:hj} we set up our notation, review the Hamilton-Jacobi
formalism, and discuss briefly the relevant features of the region of the T-models phase portrait close to the origin,
illustrating in particular the non-inflationary region $R_1$, the inflationary region $R_2$,
and the slow-roll inflationary region $R_3$, and showing that solutions with initial conditions on the plateau
of the potential eventually enter the SR region.
Section~\ref{sec:ae} is devoted to the derivation of the four asymptotic solutions (Hubble parameter and 
number of e-folds in the SR and in the KD stages) that we need for our applications. Here we give the specific
choice of $F(\Phi)$ that adapts the SR expansion for T-models. Section~\ref{sec:match} is
devoted to the matching of the asymptotic expansions for the Hubble parameter and to the determination
of the relation between the asymptotic expansions for the number of e-folds. In this section we also compare
our results with the first order approximation to the Hubble parameter in the KD stage
derived by Chowdhury, Martin, Ringeval and Vennin~\cite{CM19} using the number of e-folds as the
independent variable. In section~\ref{sec:app} we present two applications of our results: First, we use the method of matched
asymptotic expansions to compute the total amount of inflation as a function of the initial data
(which, conversely, allows us to select initial data corresponding to a fixed number of e-folds);
and second, we use our SR results to compute consistently the expansions of the spectral index $n_s(N)$
to order $1/N^2$ and of the tensor-to-scalar ratio $r(N)$ accurate to order $1/N^3$ in the SR approximation.
More precisely, we show that for T-models,
\begin{equation}
	n_s = 1 - \frac{2}{N}
	            - \frac{2}{3}\frac{\log N}{N^2}
	            + \frac{n_{s,2}(\alpha,m)}{N^2}
	            + O\left(\frac{(\log N)^2}{N^3}\right),
\end{equation}
and
\begin{equation}
	r = \frac{12\alpha}{N^2}
	     + 8\alpha\frac{\log N}{N^3}
	     + \frac{r_{2}(\alpha,m)}{N^3}
	     + O\left(\frac{(\log N)^2}{N^4}\right),
\end{equation}
where $n_{s,2}(\alpha,m)$ and $r_2(\alpha,m)$ are rather involved functions (which we compute)
of the model parameters. Note in particular the presence of intermediate logarithmic terms
between the two the standard SR approximations~\cite{KL13,KL13b,KALL15,GA15,CA15,AK18,AC21} 
and the corrections of order $1/N^2$ for $n_s(N)$ or $1/N^3$ for $r(N)$ given, for example,
in Refs.~\cite{KA13,OO16,GE21}, where these logarithmic terms are missing. Note that these are purely
SR results (without any use of the KD expansions). We also present illustrative numerical comparisons
of the accuracy of the first- and adapted second-order (purely) SR approximations as functions
of the number of e-folds and of the parameters of the model. After a summary of
our results and the aforementioned Appendix~\ref{ap:a}, we include also a brief Appendix~\ref{ap:b}
discussing the specific implementation of Pad\'e approximants used in this paper.
\section{The Hamilton-Jacobi formalism\label{sec:hj}}
In this section we review in some detail the Hamilton-Jacobi formalism of inflationary models.
Although our formulation is fairly general, from the very beginning we use a T-model as our main
example, both because of its physical relevance and because it permits us to
show all the details of the method used to adapt the slow-roll expansion to the analytic form
of the potential. In Appendix~\ref{ap:a} we show how to perform this process and give the
first few terms of the resulting expansions for the families of potentials mentioned earlier.
\subsection{Scaled magnitudes}
Hereafter we will use the reduced inflaton field $\varphi$ defined by
\begin{equation}
	\label{redinf}
	\varphi = \sqrt{\frac{3}{2}}\frac{\Phi}{{\mathrm{M}_\mathrm{Pl}}},
\end{equation}
the reduced Hubble parameter $h$ defined by
\begin{equation}
	\label{redh}
	h = 3 H,
\end{equation}
and the reduced potential $v(\varphi)$ defined by
\begin{equation}
	\label{redv}
	v(\varphi) = \frac{3}{{\mathrm{M}_\mathrm{Pl}}^2}V(\Phi) = A \left(\tanh^2(\lambda\varphi)\right)^m,
\end{equation}
where
\begin{equation}
    \label{redal}
    A = \frac{3\Lambda}{{\mathrm{M}_\mathrm{Pl}}^2},
    \quad
    \lambda = \frac{1}{3\sqrt{\alpha}}.    
\end{equation}
(The constant $A$ could be absorbed by a rescaling of the cosmic time $t$, but it is customary not to do so.)
In terms of these reduced variables, eqs.~(\ref{eq1}) and~(\ref{eq2}) read
\begin{equation}
    \label{eq:me}
    \ddot{\varphi} + h\dot{\varphi} + \frac{1}{2}v'(\varphi)= 0,
\end{equation}
and
\begin{equation}
    \label{eq:htau}
    h^2= \dot{\varphi}^2+ v(\varphi),
\end{equation}
respectively. Finally, and for later reference, we mention a useful consequence of eqs.~(\ref{eq:me}) and~(\ref{eq:htau}):
by taking the derivative of eq.~(\ref{eq:htau}) with respect to cosmic time
and eliminating $\ddot{\varphi}$ between this derivative and eq.~(\ref{eq:me}), we find that
\begin{equation}
	\label{dh}
	\dot{h} = - \dot{\varphi}^2.
\end{equation}
\subsection{The Hamilton-Jacobi formalism}
The main goal of the Hamilton-Jacobi formalism is to determine the reduced Hubble parameter $h$ as a function of the
reduced inflaton $\varphi$, i.e., to find functions $h(\varphi)$ in suitable regions of the $(\varphi,\dot{\varphi})$ phase space.
Since this is only possible in regions where $\dot{\varphi}$ has a constant sign, we first restrict our
study to the regions
\begin{equation}
	\label{D}
	D = \left\{(\varphi,\dot{\varphi}): \varphi\geq 0,\,\dot{\varphi}<0\right\},
\end{equation}
in the $(\varphi,\dot{\varphi})$ plane, and
\begin{equation}
	\label{R}
	R = \{(\varphi,h): \varphi\geq 0, \,\sqrt{v(\varphi)}<h<+\infty\},
\end{equation}
in the $(\varphi,h)$ plane,
which are related via the positive square root of eq.~(\ref{eq:htau}),
or, more formally, by the diffeomorphism $\Gamma: D \mapsto R$,
\begin{equation}
	\label{Gamma}
	\Gamma(\varphi,\dot{\varphi}) = (\varphi, \sqrt{\dot{\varphi}^2+v(\varphi)}).
\end{equation}
Note that $R$ will play the role of the phase space in the Hamilton-Jacobi formalism.

Equation~(\ref{dh}) shows that each part of a solution $\varphi(t)$ lying on $D$ satisfies
\begin{equation}
	\label{hj2}
	\dot{\varphi} = - h'(\varphi),
\end{equation}
which substituted into eq.~(\ref{eq:htau}) yields
\begin{equation}
	\label{hj1}
	h'(\varphi)^2 = h(\varphi)^2-v(\varphi).
\end{equation}
Equations~(\ref{hj2}) and~(\ref{hj1}) are referred to as the Hamilton-Jacobi formalism of 
inflationary models~\cite{SA90,LID94,HAN14,HAN19,HER19,AM20,ME20,MM21,BA09,LL09}.
A few comments are in order. First, note that eq.~(\ref{hj1}) allows us to achieve the main goal of the
formalism, i.e., to find $h(\varphi)$. Second, by integrating eq.~(\ref{hj2}), each solution $h(\varphi)$ of eq.~(\ref{hj1})
determines a corresponding solution $\varphi(t)$ in implicit form,
\begin{equation}
    \label{imp}
    t=-\int_{\varphi(0)}^{\varphi(t)}\frac{d \varphi}{h'(\varphi)}.
\end{equation}
Third, the scale factor can also be determined as a function of $\varphi$, since
\begin{equation}
	3 h'(\varphi) a'(\varphi) + h(\varphi) a(\varphi) = 0,
\end{equation}
and therefore the number of e-folds $N$ can be also determined as a function of $\varphi$ via
\begin{equation}
	\label{hj3}
	N'(\varphi) = - \frac{a'(\varphi)}{a(\varphi)} = \frac{h(\varphi)}{3 h'(\varphi)}.
\end{equation}
Finally, the symmetry
\begin{equation}
	\label{eq:sym}
	(v(\varphi),h(\varphi),\varphi(t))\rightarrow(v(-\varphi),h(-\varphi),-\varphi(t))
\end{equation}
of eqs.~(\ref{eq:me}),~(\ref{eq:htau}),~(\ref{dh}),~(\ref{hj2}) and~(\ref{hj1}) allows us to transfer the results
obtained in $D$ and $R$ to $\hat{D} = \left\{(\varphi,\dot{\varphi}):\varphi\leq 0,\,\dot{\varphi}>0\right\}$
and $\hat{R} = \{(\varphi,h): \varphi\leq 0,\,\sqrt{v(\varphi)}<h<+\infty\}$, thereby eliminating our
initial restriction.
\subsection{Phase portrait: separatrices, slow-roll and kinetic dominance\label{sec:pp}}
\begin{figure}[tbp]
	\centering
        \includegraphics[width=10cm]{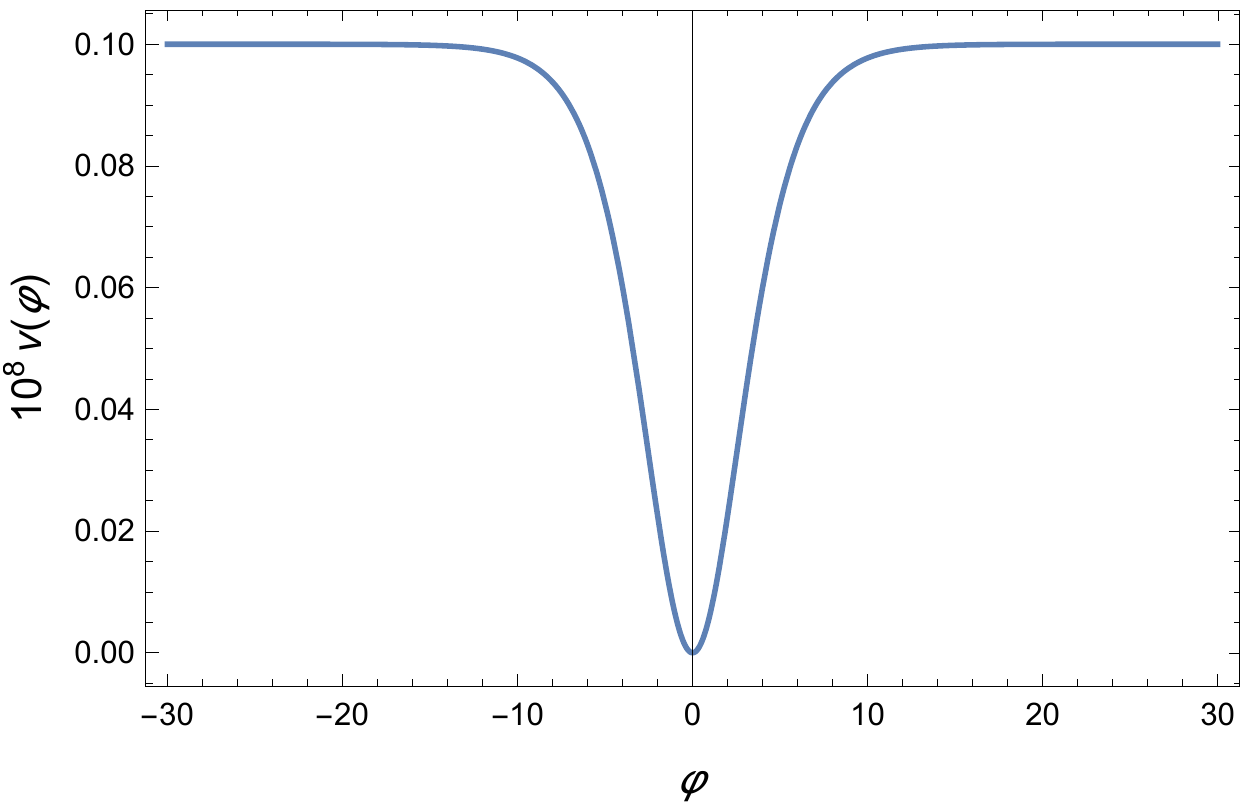}
	\caption{Graph of the reduced potential $v(\varphi) = A \left(\tanh^2(\lambda\varphi)\right)^m$ for
                  $A = 10^{-9}$, $\lambda = 1/\sqrt{15}$ and $m=1$ (parameters taken from Ref.~\cite{AK18}).
	\label{fig:v}}
\end{figure}

Figure~\ref{fig:v} shows the graph of the reduced potential eq.~(\ref{redv}) for one of the examples discussed in Ref.~\cite{AK18},
namely $A = 10^{-9}$, $\lambda = 1/\sqrt{15}$ and $m=1$, where the T-shape that gives name to these potentials and the
two concave plateaus are apparent. Note that the minimum value of the potential is $v_\mathrm{min} = v(0) = 0$.

\begin{figure}[tbp]
	\centering
        \includegraphics[width=7cm]{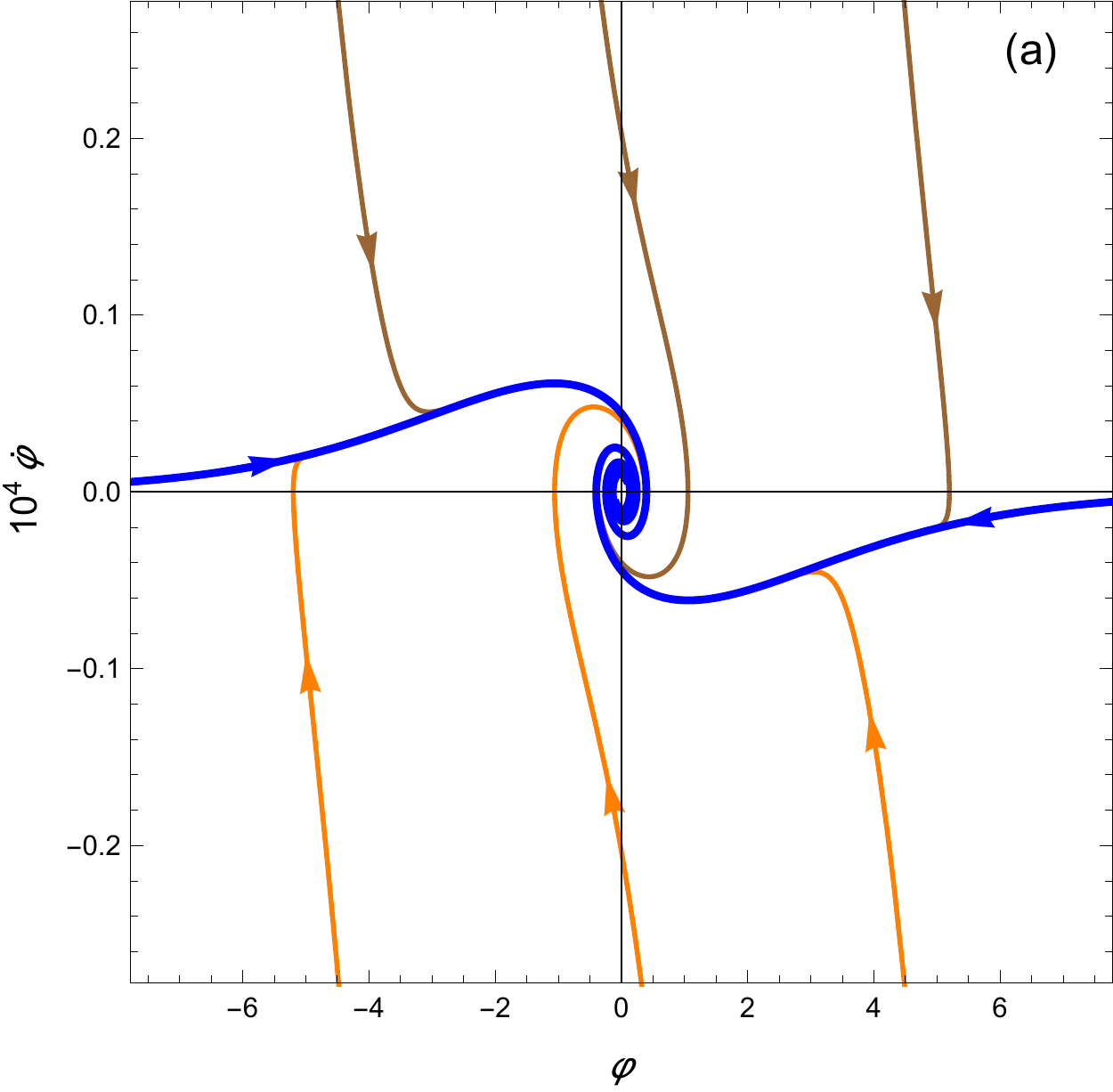}
        \hfill
        \includegraphics[width=7cm]{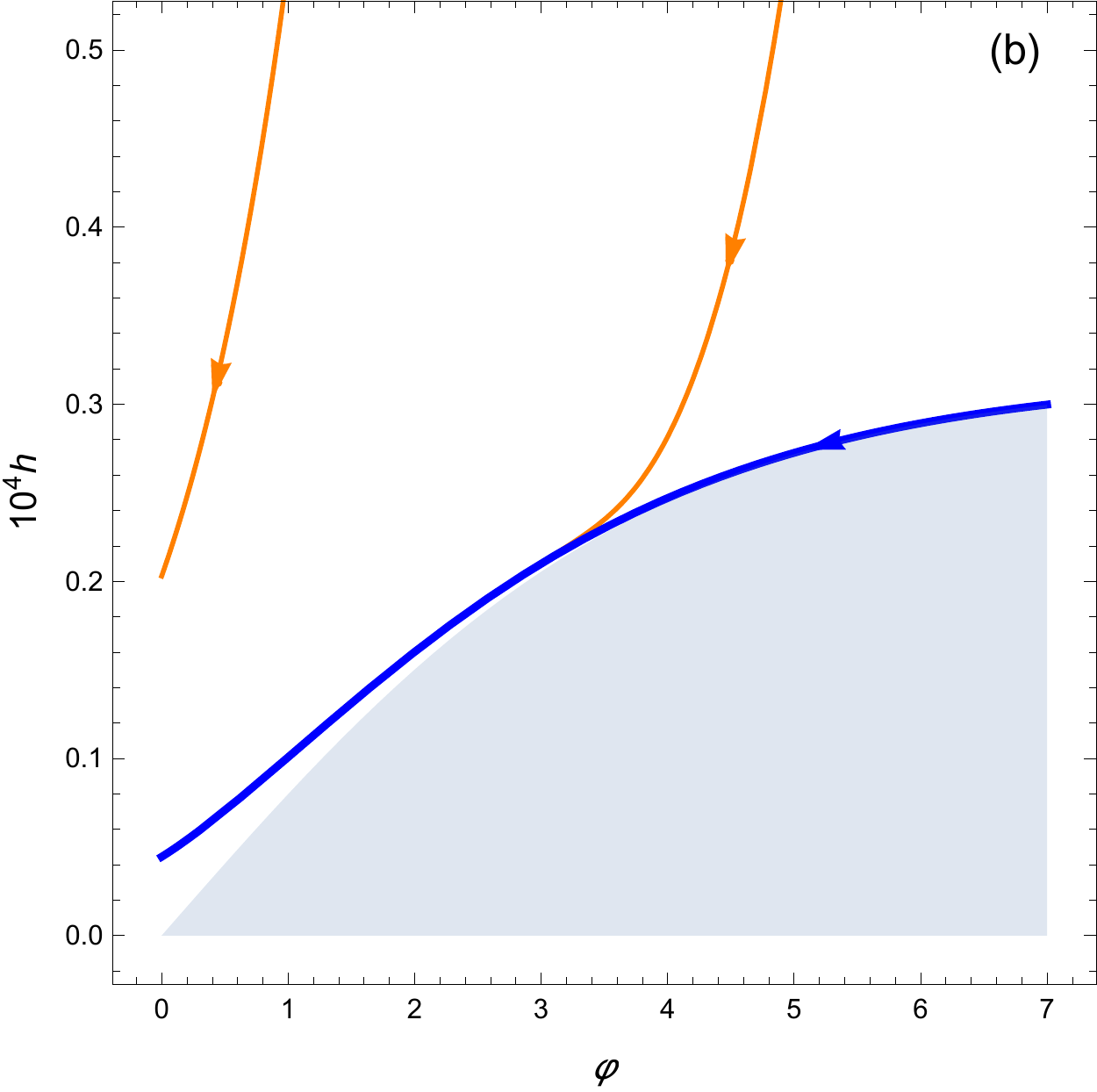}
    \caption{(a) Phase portrait in the $(\varphi,\dot{\varphi})$ plane for the potential
                       $v(\varphi) = A \left(\tanh^2(\lambda\varphi)\right)^m$ for $A = 10^{-9}$, $\lambda = 1/\sqrt{15}$ and $m=1$.
                       The two blue trajectories are the separatrices. The region $D$ defined in eq.~(\ref{D}) is the fourth quadrant.
                  (b) Corresponding phase portrait in the region $R$ defined by eq.~(\ref{R}) of the Hamilton-Jacobi $(\varphi,h)$ plane.
                        The shaded area is the forbidden region where $h\leq \sqrt{v(\varphi)}$.
	\label{fig:pp}}
\end{figure}

Figure~\ref{fig:pp}(a) shows the phase portrait for this T-model in the $(\varphi,\dot{\varphi})$ plane, and figure~\ref{fig:pp}(b)
the corresponding phase portrait in the region $R$ defined by eq.~(\ref{R}) of the Hamilton-Jacobi $(\varphi,h)$ plane, which
we now discuss briefly.

In Ref.~\cite{AM20} we proved that, for a certain class of potentials, if
\begin{equation}
	\label{condsepp}
	\lim_{\varphi\rightarrow\infty}
	\frac{v'(\varphi)}{2v(\varphi)}
	=
	\ell\quad\mbox{with} \quad 0 \leq \ell < 1,
\end{equation}
then eq.~(\ref{hj1}) has a unique solution $h_s(\varphi)$ satisfying
\begin{equation}
	\label{sepasy}
	h_s(\varphi)
	\sim
	\frac{\sqrt{v(\varphi)}}{\sqrt{1-\ell^2}}\quad\mbox{as}\quad \varphi \rightarrow \infty.
\end{equation}
The T-models given by eq.~(\ref{redv}) belong to this class of potentials with $\ell =0$.
Therefore, eq.~(\ref{hj1}) has a unique solution with asymptotic behavior
\begin{equation}
	\label{sepasyst}
	h_s(\varphi) \sim \sqrt{v(\varphi)} \quad \mbox{as} \quad \varphi \rightarrow \infty.
\end{equation}
(Incidentally, taking $h_s(\varphi) \approx \sqrt{v(\varphi)}$ is the standard SR approximation.)
This solution, colored blue in figure~\ref{fig:pp}(b), is the boundary in the region $R$ of the Hamilton-Jacobi
phase space between solutions of eq.~(\ref{hj1}) defined for all $\varphi>0$ and solutions of eq.~(\ref{hj1})
that leave $R$ at a certain $\varphi_\mathrm{m}>0$. The corresponding (full) trajectory in the $(\varphi,\dot{\varphi})$
phase plane, also colored blue in figure~\ref{fig:pp}(a), spirals in towards the origin and is part of the boundary
between regions filled by trajectories that come from large, positive values of $\dot{\varphi}$ (colored brown)
and large in magnitude, negative values of $\dot{\varphi}$ (colored orange). The remaining part of the boundary
between these trajectories is the symmetric solution. These special solutions are referred to as separatrices,
and for wide ranges of initial conditions any solution tends asymptotically to them~\cite{BE85,LID94}.

\begin{figure}[tbp]
	\centering
        \includegraphics[width=7cm]{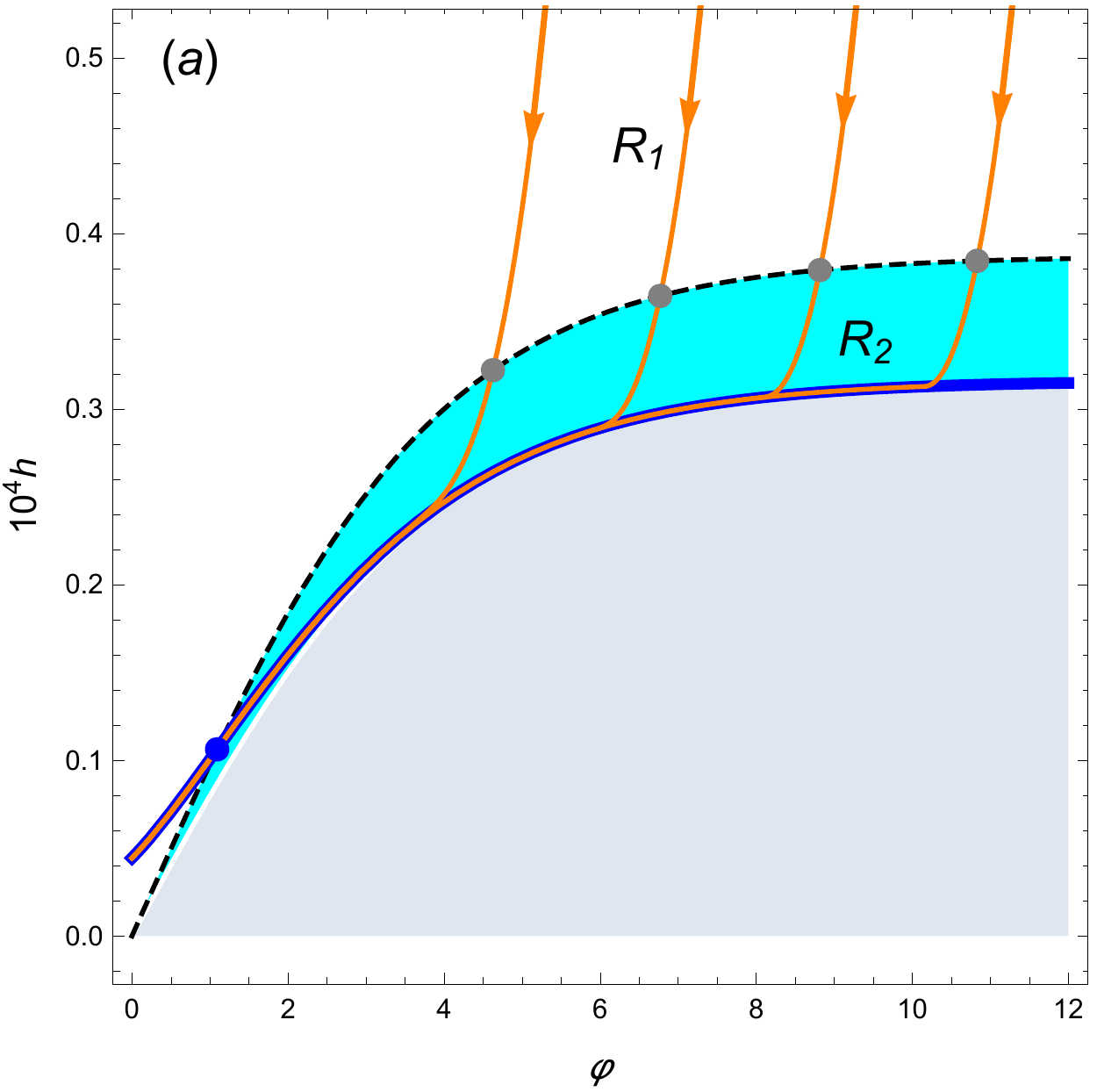}
        \hfill
        \includegraphics[width=7.25cm]{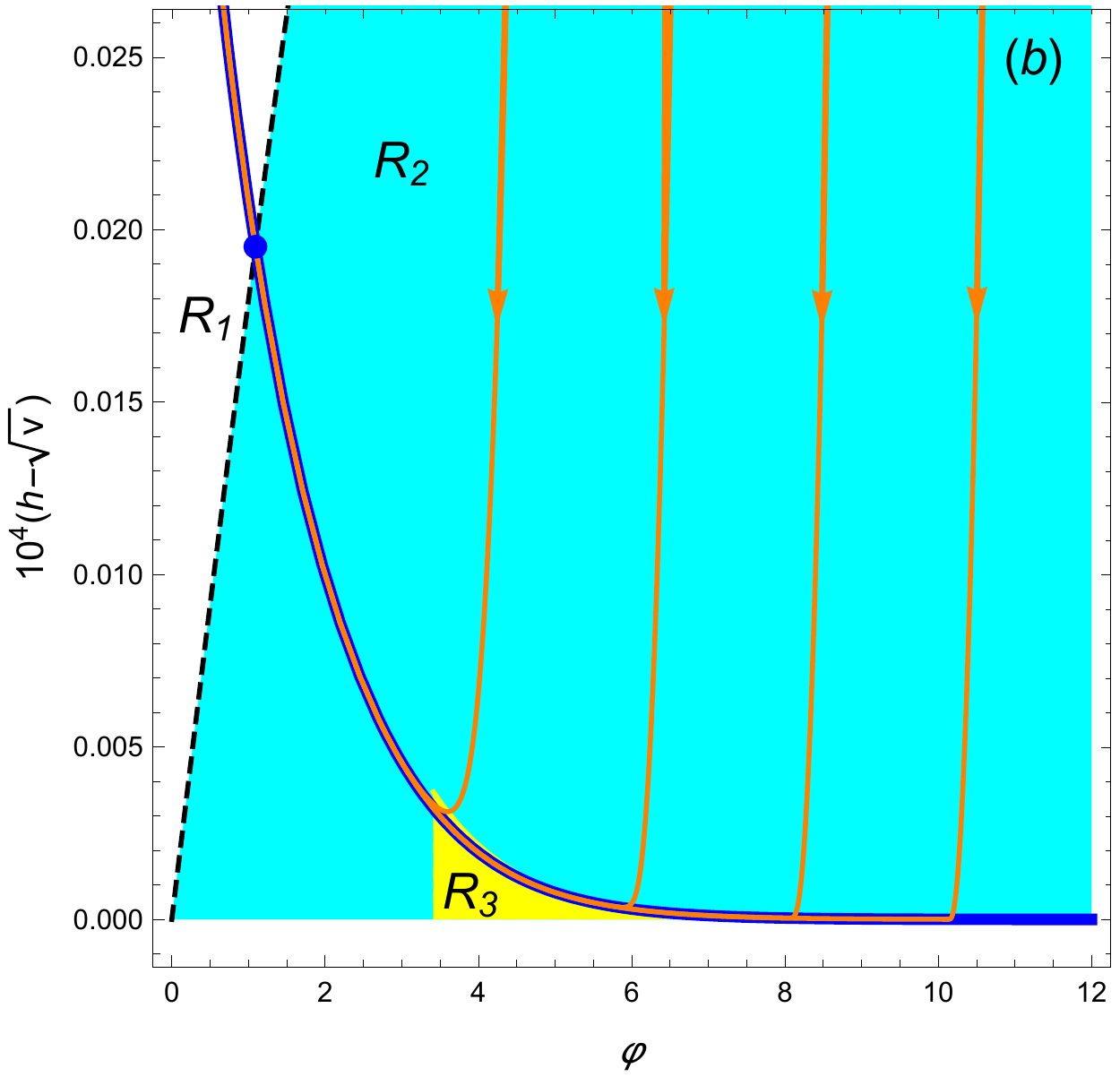}
    \caption{(a) Phase portrait in the region $R$ of the Hamilton-Jacobi $(\varphi,h)$ plane for the potential
                  $v(\varphi) = A \left(\tanh^2(\lambda\varphi)\right)^m$ for $A = 10^{-9}$, $\lambda = 1/\sqrt{15}$ and $m=1$.
                  As in figure~\ref{fig:pp}, the area shaded in gray is the forbidden region $h\leq\sqrt{v(\varphi)}$,
                  and the blue curve is the separatrix.
                  The dashed line is the curve $h(\varphi) = \sqrt{3v(\varphi)/2}$ that separates the non-inflationary region $R_1$
                  (white background) from the inflationary region $R_2$ (light blue background).
                  The gray dots mark the different values $\varphi_\mathrm{in}$ at which each trajectory enters the
                  inflationary region $R_2$, while the blue dot marks the \emph{common} value $\varphi_\mathrm{end}$
                  at which the trajectories leave the inflationary region, after being drawn towards the separatrix which,
                  although not an attractor in the strict mathematical sense (cf.~Ref~\cite{AM20}), effectively works in a similar way.
                  (b) Modified phase portrait $(\varphi, h - \sqrt{v(\varphi)})$ to show the SR region $R_3$ corresponding
                  to a value of $\varepsilon = 1/10$ in eq.~(\ref{eq:r3}).
	\label{fig:newfig}}
\end{figure}

In figure~\ref{fig:newfig} we show an enlarged portion of the region $R$ in the Hamilton-Jacobi $(\varphi,h)$ plane.
From the Hamilton-Jacobi eqs.~(\ref{hj2}) and~(\ref{hj1}) it follows that
\begin{equation}
   \ddot{a} = \frac{a}{3}\left(v(\varphi)-\frac{2}{3}h(\varphi)^2\right),
\end{equation}
and therefore the inflation region $\ddot{a}>0$ corresponds to
\begin{equation}
	\label{eq:ir}
	\sqrt{v(\varphi)} < h(\varphi) < \sqrt{\frac{3}{2} v(\varphi)}
\end{equation}
(the lower bound is just eq.~(\ref{R})). As in figure~\ref{fig:pp}, the area shaded in gray in figure~\ref{fig:newfig}(a)
is the forbidden region $h\leq\sqrt{v(\varphi)}$, and the blue curve is the separatrix.
The dashed line is the curve $h(\varphi) = \sqrt{3v(\varphi)/2}$ that separates the non-inflationary region $R_1$
(white background) from the inflationary region $R_2$ (light blue shading). Note that not all trajectories enter
the inflationary region $R_2$ (for example, the leftmost trajectory in figure~\ref{fig:pp}(b)), but that
all trajectories with initial conditions above the plateau eventually enter $R_2$.
The gray dots mark the different values $\varphi_\mathrm{in}$ at which these trajectories enter the
inflationary region $R_2$, while the blue dot marks the \emph{common} value $\varphi_\mathrm{end}$
at which the trajectories leave the inflationary region, after being drawn towards the separatrix.
The SR stage is typically defined by,
\begin{equation}
	\label{eq:srcrit}
	\left| \epsilon_{1,V} \right| =  \frac{3}{4} \left| \frac{v'(\varphi)}{v(\varphi)} \right|^2 < \varepsilon,
	\quad
	\left| \epsilon_{2,V} \right| =   \frac{3}{2} \left| \frac{v''(\varphi)}{v(\varphi)} \right| < \varepsilon,	
\end{equation}
where $\varepsilon \ll 1$. Note that these conditions define only an interval $\varphi \geq \varphi_1$.
To define an SR region $R_3$ in the $(\varphi,h)$ plane and following the ideas that lead
to the criterion~(\ref{eq:srcrit}) we use
\begin{equation}
	h(\varphi) \approx\sqrt{v(\varphi)},
	\quad
	\dot{\varphi}\approx-\frac{v'(\varphi)}{2h}\approx-\frac{v'(\varphi)}{2\sqrt{v(\varphi)}},
\end{equation}
to obtain the improved approximation,
\begin{equation}
	h(\varphi) \approx\sqrt{\frac{v'(\varphi)^2}{4{v(\varphi)}}+v(\varphi)},
\end{equation}
which allows as to define the ($\varepsilon$-dependent, through $\varphi_1$) SR region as
\begin{equation}
	\label{eq:r3}
	R_3 =
	\left\{ (\varphi,h):\varphi>\varphi_1\quad\mbox{and}\quad \sqrt{v(\varphi)}\leq h \leq \sqrt{\frac{v'(\varphi)^2}{4{v(\varphi)}}+v(\varphi)}\right\}.
\end{equation}
As an illustration, in figure~\ref{fig:newfig}(b) we show a modified phase portrait in which,
instead of $(\varphi, h)$ we plot $(\varphi, h-\sqrt{\varphi})$ for a value of $\varepsilon=1/10$.
Note that the $R_3$ region extends indefinitely to the right, that the separatrix provides an
accurate approximations to the solutions of the T-model in the SR stage, and that $\varphi_\text{end}$
lies outside $R_3$.

Let us consider now the behavior of the solutions backwards in the cosmic time $t$. Equation~(\ref{dh}) shows
that the reduced Hubble parameter $h$ is a positive, monotonically decreasing function of $t$.
Therefore, the reduced Hubble parameter $h(t)$ increases backwards in time and both
$h(t)$ and $\varphi(t)$ may develop singularities. In the KD stage, where
\begin{equation}
    \label{kd}
    \dot{\varphi}^2 \gg v(\varphi),
\end{equation}
we may neglect $v(\varphi)$ in eq.~(\ref{hj1}) and obtain the approximate equation
\begin{equation}
    \label{appr}
    h'(\varphi) \sim \pm h(\varphi),
\end{equation}
which yields  two families of  approximate solutions
\begin{equation}
	\label{hvarphiasyp}
	h(\varphi) \sim\frac{e^{\varphi}}{b}\quad \mbox{as} \quad \varphi\rightarrow\infty,
\end{equation}
and
\begin{equation}
	\label{hvarphiasym}
	h(\varphi) \sim\frac{e^{-\varphi}}{b}\quad \mbox{as} \quad \varphi\rightarrow-\infty,
\end{equation}
where $b$ is a strictly positive but otherwise arbitrary parameter. Incidentally, note that
the solution to the initial value problem derived from eqs.~(\ref{hj2}) and~(\ref{hvarphiasyp}), 
\begin{equation}
	\label{eq:dekd}
	\dot{\varphi}(t) \approx - \frac{e^{\varphi(t)}}{b},
	\quad
	\varphi(t_0) = \varphi_0,
\end{equation}
is given by
\begin{equation}
	\varphi(t) \approx \varphi_0 - \log\left( 1 + \frac{e^{\varphi_0}}{b} (t-t_0)\right),
\end{equation}
and
\begin{equation}
	\dot{\varphi}(t) = - \frac{e^{\varphi_0}}{b} \left( 1 + \frac{e^{\varphi_0}}{b} (t-t_0)\right)^{-1},
\end{equation}
which corresponds to eqs.~(4.4a,b) derived by Goldwirth and Piran~\cite{GP92} for their case $\tilde{\Pi}_0 <0$,
while the analogous result for their case $\tilde{\Pi}_0>0$,
\begin{equation}
	\dot{\varphi}(t) = \frac{e^{-\varphi_0}}{b} \left( 1 + \frac{e^{-\varphi_0}}{b} (t-t_0)\right)^{-1},
\end{equation}
can be derived from our eqs.~(\ref{hj2}) and~(\ref{hvarphiasym}).

Again, due to the symmetry eq.~(\ref{eq:sym}), we can  restrict our analysis to solutions with the asymptotic
behavior given by eq.~(\ref{hvarphiasyp}), for which the integral in the right-hand side
of Eq~(\ref{imp}) converges as $\varphi\rightarrow\infty$, i.e., these solutions emerge
from the KD stage and blow up at a finite time
\begin{equation}
    \label{imps}
    t^*=-\int_{\varphi(0)}^{\infty}\frac{d \varphi}{h'(\varphi)}.
\end{equation}
These singularities, however, lie outside the domain where the KD asymptotic expansions derived
in the following section are valid~\cite{AM22}.
\section{Asymptotic expansions\label{sec:ae}}
In this section we derive asymptotic expansions for the reduced Hubble parameter $h(\varphi)$ and the number of e-folds
$N(\varphi)$ both in the SR and in the KD stages. Note that 
\begin{equation}
	\mathcal{V}(\Phi) = \sqrt{\frac{2}{3\alpha}} \frac{m}{\mathrm{M}_\mathrm{Pl}}
	\text{csch}\left(\sqrt{\frac{2}{3\alpha}} \frac{\Phi}{\mathrm{M}_\mathrm{Pl}}\right),
\end{equation}
and that
\begin{equation}
	\text{csch}\,x = 2 \sum_{n=0}^\infty \frac{1}{e^{(2n+1)x}},
	\quad
	\mbox{as $x\to\infty$}.
\end{equation}
Therefore we adapt these expansions to the form of the potential~(\ref{Tmodel}) by taking the function $F(\Phi)$
in eq.~(\ref{eq:qf}) as,
\begin{equation}
	F(\Phi) = \exp\left(\sqrt{\frac{2}{3\alpha}} \frac{\Phi}{\mathrm{M}_\mathrm{Pl}}\right),
\end{equation}
or, in scaled variables,
\begin{equation}
	f(\varphi) = e^{2\lambda\varphi},
\end{equation}
and we take as our new independent variable
\begin{equation}
	\label{eq:y}
	y = \frac{1}{f(\varphi)} = e^{-2\lambda\varphi},
\end{equation}
with the corresponding new functions
\begin{equation}
	\label{eq:hhat}
	\hat{h}(y) = h(\varphi),
\end{equation}
\begin{equation}
	\label{eq:ahat}
	\hat{N}(y) = N(\varphi).
\end{equation}
Substituting eqs.~(\ref{eq:y}) and~(\ref{eq:hhat}) into eq.~(\ref{hj1}) with the potential given by eq.~(\ref{redv}),
we find the following equation for $\hat{h}(y)$,
\begin{equation}
	\label{eq:hhatde}
	4 \lambda^2 y^2 \hat{h}'(y)^2 - \hat{h}(y)^2 + A \left(\frac{1-y}{1+y}\right)^{2m} = 0,
\end{equation}
and substituting into eq.~(\ref{hj3}), the following equation for $\hat{N}(y)$,
\begin{equation}
	\label{eq:ahatde}
	\hat{N}'(y) = \frac{1}{12\lambda^2 y^2} \frac{\hat{h}(y)}{\hat{h}'(y)}.
\end{equation}
\subsection{The Hubble parameter in the SR stage}
As we discussed in section~\ref{sec:pp}, the separatrix is an accurate approximation to the solutions
in the SR stage, and since the separatrix is uniquely identified by the asymptotic behavior given in eq.~(\ref{sepasyst}),
we look for an asymptotic solution of eq.~(\ref{eq:hhatde}) in the form of the leading asymptotic prefactor times a power series in $y$,
\begin{equation}
	\label{eq:hhatfs}
	\hat{h}_\text{SR}(y) = \sqrt{A} \left(\frac{1-y}{1+y}\right)^{m}
	                      \left( 1 +  8 m^2 \lambda^2 \sum_{n=2}^\infty (-1)^n c_n y^n \right).
\end{equation}
By substituting this ansatz into eq.~(\ref{eq:hhatde}) we find that $c_2=1$ and that the unknown coefficients $c_n$
(which depend on $\lambda$ and $m$) can be computed from the recurrence relation
\begin{eqnarray}
   \label{cs}
   c_n & = & 8 m\lambda^2(n-1)c_{n-1}+2(1+8m^2\lambda^2)c_{n-2}-8m\lambda^2(n-3)c_{n-3}-c_{n-4} \nonumber \\
          &    & {}-4m^2\lambda^2\Big(\sum_{j=2}^{n-6}c_jc_{n-4-j}-2\sum_{j=2}^{n-4}c_jc_{n-2-j}+\sum_{j=2}^{n-2}c_jc_{n-j}\Big) \nonumber\\
          &    & {}+16m^2\lambda^4\sum_{j=1}^{n-3}
                   [(j+1)c_{j+1}+2mc_j-(j-1)c_{j-1}]\nonumber\\
          &   & \qquad\qquad\qquad\times        [(n-j-1)c_{n-j-1}+2mc_{n-j-2}-(n-j-3)c_{n-j-3}],
\end{eqnarray}
where we have set $c_n=0$ for $n<2$. The resulting even and odd coefficients can be written as,
\begin{equation}
	\label{eq:ceven}
	c_{2j} = \sum_{k=0}^{j-1} (m \lambda)^{2k} p_{2j}^{(k)} (\lambda^2),
\end{equation}
\begin{equation}
	\label{eq:codd}
	c_{2j+1} = m\lambda^2 \sum_{k=0}^{j-1} (m \lambda)^{2k} p_{2j+1}^{(k)} (\lambda^2),
\end{equation}
respectively, where the $p_{n}^{(k)} (\lambda^2)$ are polynomials of degree $k$ in $\lambda^2$ which we list in Table~\ref{tab:cn}
up to $n=9$.
\begin{table}[tbp]
\centering
\begin{tabular}{ccl}
\hline
  $n$ & $k$ & $p_n^{(k)}(\lambda^2)$ \\
\hline
  $2$ & $0$ & $1$   \\
  $3$ & $0$ & $16$ \\
  $4$ & $0$ & $2$ \\
         & $1$ & $12 + 448 \lambda^2 $ \\
  $5$ & $0$ & $80$ \\
         & $1$ & $640 + 17408 \lambda^2 $ \\
  $6$ & $0$ & $3$ \\
         & $1$ & $48 + 4096 \lambda^2 $ \\
         & $2$ & $160 + 36352 \lambda^2 + 847872 \lambda^4$ \\
  $7$ & $0$ & $224$ \\
         & $1$ & $4480 + 248832 \lambda^2$ \\
         & $2$ & $17920 + 2326528 \lambda^2+ 49020928 \lambda^4$ \\
  $8$ & $0$ & $4$ \\
         & $1$ & $120 + 18304 \lambda^2$ \\
         & $2$ & $960 + 392192 \lambda^2+ 17276928 \lambda^4$ \\
         & $3$ & $2240 + 1714176 \lambda^2+ 166674432 \lambda^4 + 3255828480 \lambda^6$ \\
  $9$ & $0$ & $480$ \\
         & $1$ & $17280 + 1612800 \lambda^2$ \\
         & $2$ & $161280 + 35618816 \lambda^2+ 1340866560 \lambda^4$ \\
         & $3$ & $430080 + 163872768 \lambda^2+ 13211009024 \lambda^4 + 243139608576 \lambda^6$\\
\hline
\end{tabular}
\caption{\label{tab:cn}Polynomials $p_n^{(k)}(\lambda^2)$ that give the coefficients $c_n$ (up to $n=9$)
              in the formal expansion of the separatrix $\hat{h}_\text{SR}(y)$ as functions of the parameters $m$ and $\lambda$
              for the T-models with reduced potentials $v(\varphi) = A (\tanh^2(\lambda\varphi))^m$.}
\end{table}
\subsection{The number of e-folds in the SR stage\label{sec:neSR}}
The corresponding asymptotic expansion for the number of e-folds in the SR stage follows immediately from eq.~(\ref{eq:ahatde}),
\begin{equation}
	\label{eq:ahatsy}
	\hat{N}_\text{SR}(y) - \hat{N}_{\text{SR},0} = \frac{1}{12\lambda^2} \int \frac{\hat{h}_\text{SR}(y)}{\hat{h}_\text{SR}'(y)}\frac{dy}{y^2}.
\end{equation}
After term by term integration and separation of the leading terms we get an asymptotic expansion which we write in the form
\begin{equation}
	\label{eq:aSR}
	\hat{N}_\text{SR}(y) - \hat{N}_{\text{SR},0} = \frac{1}{24 m \lambda^2}
	                                                                          \left(y+\frac{1}{y}\right) - \frac{1}{3} \log y + \sum_{n=1}^\infty (-1)^{n+1} \gamma_n y^n.
\end{equation}
Note that there are two terms in $y$ in this expansion: one in the term between parentheses and one in
the formal series. The reason for this choice is that (aside from the dependence of the integration constant) the term
in parentheses is the SR approximation eq.~(\ref{standardSRPhi}): Indeed, if we integrate eq.~(\ref{hj3})
with the condition $N(\varphi_\mathrm{end}) = 0$,
\begin{equation}
	N(\varphi) = \frac{1}{3} \int_{\varphi_\mathrm{end}}^\varphi\frac{h(s)}{h'(s)}\, ds,
\end{equation}
and use the SR approximation (i.e., the leading term of the SR expansion) $h(s)\approx \sqrt{v(s)}$, we find that
\begin{equation}
	N_\mathrm{SR,0}(\varphi) = \frac{2}{3} \int_{\varphi_\mathrm{end}}^\varphi\frac{v(s)}{v'(s)}\, ds,
\end{equation}
or, using the reduced form of the potential for T-models~(\ref{redv}),
\begin{equation}
	N_\mathrm{SR,0}(\varphi) = \frac{1}{12m\lambda^2} \left(\cosh(2\lambda\varphi) - \sqrt{12\lambda^2m^2+1}\right).
\end{equation}
This is eq.~(\ref{standardSRPhi}) in reduced variables, and eq.~(\ref{eq:y}) leads to
\begin{equation}
	\label{eq:nnyy}
	\hat{N}_\mathrm{SR}(y) - \hat{N}_\mathrm{SR,0} = \frac{1}{24m\lambda^2}\left(y+\frac{1}{y}\right),
\end{equation}
which is precisely the first term in the right-hand side of eq.~(\ref{eq:aSR}). Incidentally, this result shows that the
logarithmic term in eq.~(\ref{eq:aSR}) is a second-order term in the SR expansion.

The coefficients $\gamma_n$ in eq.~(\ref{eq:aSR}) can be readily calculated from the $c_2,\ldots,c_{n+2}$ during the integration.
For example,
\begin{eqnarray}
	\gamma_1 & = & \frac{16}{3} m \lambda^2, \label{eq:g1} \\
	\gamma_2 & = & \frac{1}{3} + \frac{8}{3}  m^2 \lambda^2 + 96 m^2  \lambda^4, \label{eq:g2}\\
	\gamma_3 &= & \frac{112}{9} m\lambda^2 + \frac{1024}{9} m^3\lambda^4
	                          + \frac{28672}{9}  m^3 \lambda^6, \label{eq:g3}
\end{eqnarray}
but a more efficient method to compute these coefficients is to substitute eqs.~(\ref{eq:hhatfs}) and~(\ref{eq:aSR}) into
eq.~(\ref{eq:ahatde}), obtaining directly the relation between the $\gamma_n$ and the $c_n$. Thus, we reproduce eqs.~(\ref{eq:g1})--(\ref{eq:g3})
and find that for $n\ge 4$,
\begin{eqnarray}
   \label{gammas}
   \gamma_n & = & \frac{1}{6n}
                              \Big[(n+2)c_{n+2}-nc_n+2mc_{n-1} 
                              -8m\lambda^2 \left( (n+1)c_{n+1} + 2 m c_n -  (n-1) c_{n-1} \right) \nonumber \\
                      & & {}- (1+24 m \lambda^2 \gamma_1) (n c_n + 2 m c_{n-1} - (n-2) c_{n-2})
                                - 48 m \lambda^2 (n-1) \gamma_{n-1} \nonumber \\
                      & & {}- 24 m \lambda^2 \sum_{j=3}^{n-1} \left( (n+2-j) c_{n+2-j} + 2 m c_{n+1-j} - (n-j) c_{n-j} \right) (j-1) \gamma_{j-1}\Big].
 \end{eqnarray}
\subsection{The Hubble parameter in the KD stage}
Similarly, to find an asymptotic expansion for the reduced Hubble parameter in the KD stage we have to look for a formal
solution of eq.~(\ref{eq:hhatde}) with leading asymptotic behavior given by eq.~(\ref{hvarphiasyp}). However, as we mentioned
in the Introduction, and because of this leading asymptotic behavior, this formal solution has to be of the form,
\begin{equation}
	\label{kd1}
	\hat{h}_\text{KD}(y,b) = \frac{1}{b y^{\frac{1}{2\lambda}}}
	                 + \sum_{n=1}^{\infty} A^n b^{2n-1}
	                   \zeta_n(y) y^\frac{2n-1}{2\lambda},
\end{equation}
where $\zeta_n(y)$ are in turn formal power series in $y$,
\begin{equation}
	\label{zetan}
	\zeta_n(y) = \sum_{p=0}^{\infty}(-1)^p\zeta_{n,p} y^p.
\end{equation}
For the computation of the coefficients $\zeta_{n,p}$ we assume temporarily that $\lambda$ is not a rational number,
which makes integer powers of $y$ and of $y^{\frac{1}{2\lambda}}$ linearly independent, and thus allows independent
identification of the corresponding coefficients. A limiting argument shows that our results
are still valid for rational values of $\lambda$, and a somewhat lengthy computation shows that the coefficients are determined
recursively by
\begin{eqnarray}
	\label{eq:z10}
	\zeta_{1,0} & = & \frac{1}{4}, \\
	\label{eq:z1p}
	\zeta_{1,p} & =  & \frac{1}{4(1 + p \lambda)} \sum_{j=1}^{p} 2^j \binom{2m}{j}\binom{p-1}{j-1},
	\quad (p=1,2,\ldots),
\end{eqnarray}
\begin{equation}
	\label{kd2nsol}
	\zeta_{n+1,p} = \frac{1}{n+1+p\lambda}\sum_{j=1}^n\sum_{r=0}^p
	\left[\left(j+r\lambda-\frac{1}{2}\right)\left(n-j+(p-r)\lambda+\frac{1}{2}\right)-\frac{1}{4}\right]\zeta_{j,r}\zeta_{n+1-j,p-r}.
\end{equation}
Since eq.~(\ref{kd2nsol}) for $p=0$ takes the form
\begin{equation}
	\label{pzero}
	\zeta_{n+1,0}
	=
	\frac{1}{n+1}\sum_{j=1}^n\left[\left(j-\frac{1}{2}\right)\left(n-j+\frac{1}{2}\right)-\frac{1}{4}\right]\zeta_{j,0}\zeta_{n+1-j,0},
\end{equation}
by setting $n=1$ we find that $\zeta_{2,0} = 0$, and by induction that $\zeta_{n,0}=0$ for all $n\geq 2$.
For example, the first three coefficients are,
\begin{equation}
	\zeta_{1}(y)
	=
	\frac{1}{4}
	 -  \frac{m}{\lambda +1} y
	+ \frac{2 m^2}{2 \lambda+1} y^2
	+ \cdots,
\end{equation}
\begin{equation}
	\zeta_2(y)
	=
	 - \frac{\lambda  m}{4 \left(\lambda+1\right)\left(\lambda+2\right)} y
	+ \frac{\lambda  (3 \lambda +2) m^2}{2 (\lambda +1)^2 (2 \lambda+1)} y^2
	+ \cdots,
\end{equation}
\begin{equation}
	\zeta_3(y)
	=
	 - \frac{\lambda  m}{16 (\lambda +2) (\lambda+3)} y
	+
	\frac{\lambda  \left(7 \lambda ^2+16 \lambda +6\right) m^2}{8 (\lambda +1)^2 (\lambda +2) (2 \lambda+3)} y^2
	+
	\cdots.
\end{equation}
\subsection{The number of e-folds in the KD stage}
Similarly, asymptotic solutions for the number of e-folds in the KD stage have the form,
\begin{equation}
	\label{akd1}
	\hat{N}_\text{KD}(y,b) - \hat{N}_{\text{KD},0}(b)
	=
	- \frac{\log y}{6\lambda} - \sum_{n=1}^{\infty}A^n b^{2n}\xi_n(y) y^\frac{n}{\lambda},
\end{equation}
where
\begin{equation}
	\xi_n(y) = \sum_{p=0}^{\infty}(-1)^p\xi_{n,p} y^p,
\end{equation}
and using eq.~(\ref{eq:ahatde}) we find that the coefficients $\xi_{n,p}$ can be computed recursively from
\begin{equation}
	\label{a1pKD}
	\xi_{1,p} = \frac{1}{3}\zeta_{1,p},
\end{equation}
and for $n \geq 2$
\begin{equation}
	\label{anKD}
	\xi_{n,p} = \frac{1}{3}\zeta_{n,p}
	              +
	              \frac{1}{n+p\lambda}
	              \sum_{j=1}^{n-1}\sum_{r=0}^p\left[(j+r\lambda)(2(n-j)+2(p-r)\lambda-1)\right]\xi_{j,r}\zeta_{n-j,p-r}.
\end{equation}
The first three coefficients are,
\begin{equation}
	\xi_{1}(y)
	=
	\frac{1}{12}
	-\frac{m}{3 (\lambda +1)} y
	+\frac{2 m^2}{3 (2 \lambda+1)} y^2
	+\cdots,
\end{equation}
\begin{equation}
	\xi_{2}(y)
	=
	\frac{1}{96}
	-\frac{(2\lambda+1) m}{6 (\lambda +1)(\lambda +2)} y
	+\frac{\left(5 \lambda ^2+5 \lambda+1\right) m^2}{3 (\lambda +1)^2 (2 \lambda+1)} y^2
	+
	\cdots,
\end{equation}
\begin{equation}
	\xi_{3}(y)
	=
	\frac{1}{576}
	-\frac{\left(9 \lambda ^2+19\lambda +6\right) m}{48 (\lambda +1) (\lambda+2) (\lambda +3)} y
        +\frac{\left(90 \lambda ^4+313 \lambda^3+338 \lambda ^2+141 \lambda +18\right) m^2
          }{24 (\lambda +1)^2 (\lambda +2) (2 \lambda+1) (2 \lambda +3)} y^2
   	+
	\cdots.
\end{equation}
\section{Matching of the SR and KD asymptotic expansions\label{sec:match}}
In the previous section we have found asymptotic solutions
\begin{equation}
	h_\text{SR}(\varphi) = \hat{h}_\text{SR}(e^{-2\lambda\varphi}),
	\quad
	N_\text{SR}(\varphi) = \hat{N}_\text{SR}(e^{-2\lambda\varphi}),
\end{equation}
and
\begin{equation}
	h_\text{KD}(\varphi,b) = \hat{h}_\text{KD}(e^{-2\lambda\varphi},b),
	\quad
	N_\text{KD}(\varphi,b) = \hat{N}_\text{KD}(e^{-2\lambda\varphi},b),
\end{equation}
of eqs.~(\ref{hj1}) and~(\ref{hj3}) valid, in principle, in the SR and KD stages respectively. In this section we discuss how to match
these asymptotic solutions to cover the whole inflation region, thereby allowing us to obtain approximate values of relevant
magnitudes as functions of the parameter $b$ in eq.~(\ref{hvarphiasyp}), or equivalently,
as functions of the initial conditions in eq.~(\ref{eq:dekd}),
\begin{equation}
	\label{eq:icb}
	b \approx - \frac{e^{\varphi_0}}{\dot{\varphi}_0} = \frac{e^{\varphi_0}}{\sqrt{h_0^2-v(\varphi_0)}},
\end{equation}
which in turn will allow us to find initial conditions that correspond to a previously
fixed amount of inflation. This is the layout of the procedure:
\begin{enumerate} 
	\item The $h_\text{KD}(\varphi,b)$ asymptotic expansion, when appropriately Pad\'e-summed,
                 extends its domain of validity beyond the region where eq.~(\ref{kd}) is satisfied and enters
                 the so-called ``fast roll'' stage, which allows us to compute the beginning of the inflation
                 interval~(\ref{eq:ir}) (traveled from right to left in our plots, cf.~figure~\ref{fig:newfig}) by the condition
\begin{equation}
	\label{inKD}
	h_\text{KD}(\varphi_\text{in}(b),b) = \sqrt{\frac{3}{2} v(\varphi_\text{in}(b))}.
\end{equation}
                 Note that the dependency of $\varphi_\text{in}$ on $b$ encodes the initial condition.
	\item Similarly, the $h_\text{SR}(\varphi)$ asymptotic expansion, when appropriately Pad\'e-summed,
	        extends its domain of validity beyond the SR stage to the value $\varphi_\text{end}$ where inflation ends,
\begin{equation}
	\label{endSR}
	h_\text{SR}(\varphi_\text{end}) = \sqrt{\frac{3}{2} v(\varphi_\text{end})}.
\end{equation}
	        Note that since we have approximated all the solutions in the SR stage by the
                 separatrix, this value $\varphi_\text{end}$ (reached by Pad\'e summation) will be independent of $b$.
	\item To determine an intermediate value $\varphi_*(b)$ such that $h(\varphi)$ is well
approximated by $h_\text{SR}(\varphi)$ on $[\varphi_\text{end},\varphi_*(b)]$, by $h_\text{KD}(\varphi,b)$
on $[\varphi_*(b),\varphi_\text{in}(b)]$, and by both on a neighborhood of $\varphi_*(b)$, we select
$\varphi_*(b)$ as the first local minimum of $(h_\text{KD}(\varphi,b) - h_\text{SR}(\varphi))^2$,
i.e., the local minimum closest to $\varphi_\text{in}(b)$.
	\item Next, to determine the integration constants $\hat{N}_{\text{SR},0}$ and $\hat{N}_{\text{KD},0}(b)$ in the asymptotic expansions for $N$,
we first fix the origin from which we count the amount of inflation by $N_\text{SR}(\varphi_\text{end})=0$, i.e.,
\begin{equation}
	\label{eqastarb}
	\hat{N}_\text{SR}(e^{-2\lambda\varphi_\text{end}}) = 0.
\end{equation}
	\item Finally, we determine $\hat{N}_{\text{KD},0}(b)$ by continuity at $\varphi_*(b)$, i.e., by
\begin{equation}
	\label{astarb}
	\hat{N}_\text{SR}(e^{-2\lambda\varphi_*(b)}) = \hat{N}_\text{KD}(e^{-2\lambda\varphi_*(b)},b).
\end{equation}
\end{enumerate} 

\begin{figure}[tbp]
	\centering
        \includegraphics[width=7cm]{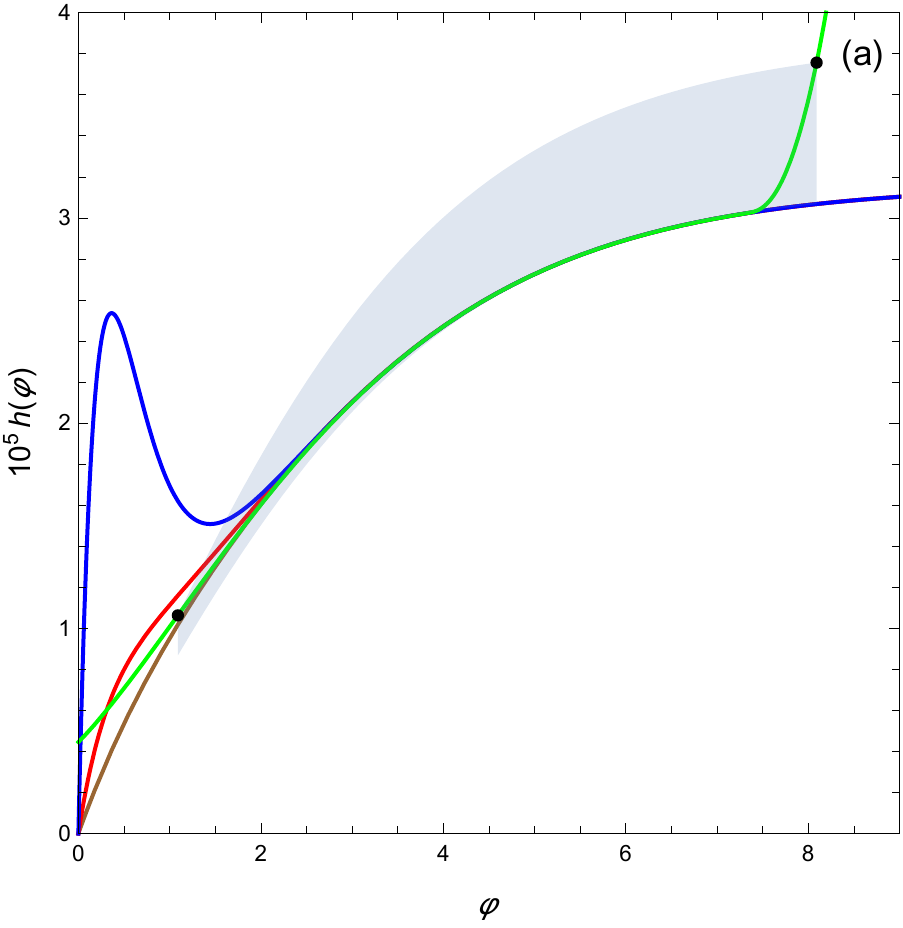}
        \hfill
        \includegraphics[width=7cm]{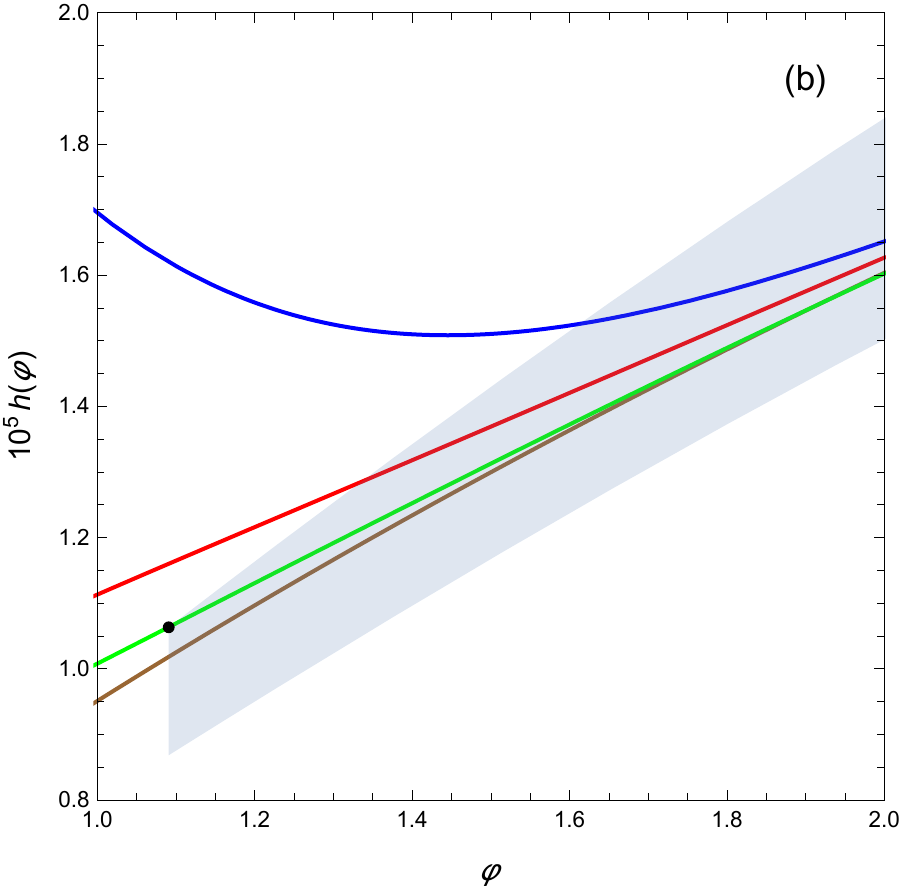}
    \caption{Numerical integration (green curve) and partial sums of $h_\text{SR}(\varphi)$ to $n_\text{SR}=2$, $4$, and
                  $6$ terms (brown, red and blue curves, respectively) for a T-model with $A = 10^{-9}$,
                  $\lambda = 1/\sqrt{15}$ and $m=1$ with initial condition  $h(\varphi_0)=2\times 10^{-4}$ at $\varphi_0=10$
                  (outside the range of the figure): (a) Inflation region as given by eq.~(\ref{eq:ir});
                  (b) Magnification around $\varphi_\text{end}\approx1.091$.
	\label{fig:srps}}
\end{figure}

Key to this matching procedure is the existence of a neighborhood of $\varphi_*(b)$
on which the appropriately summed expansions $h_\text{SR}(\varphi)$ and $h_\text{KD}(\varphi,b)$
are both accurate. We consider first the SR asymptotic expansion.
The green curve in figure~\ref{fig:srps} is the result of a numerical integration for the T-model
with $A = 10^{-9}$, $\lambda = 1/\sqrt{15}$ and $m=1$ with initial condition 
$h(\varphi_0)=2\times 10^{-4}$ at $\varphi_0=10$ (outside the range shown in the figure).
The shaded region is the inflation region $R_2$ as defined in eq.~(\ref{eq:ir}), and the
brown, red, and blue curves correspond to the partial sums of the formal series in eq.~(\ref{eq:hhatfs})
to $n_\text{SR}=2$, $4$ and $6$ terms, respectively. Figure~\ref{fig:srps}(a) shows that (in part because of the leading
behavior as $\varphi\to\infty$ built in eq.~(\ref{eq:hhatfs})) all these partial sums give excellent
approximations at large $\varphi$, but the magnification in figure~\ref{fig:srps}(b) shows that partial sums
are clearly insufficient to approximate accurately the reduced Hubble parameter in a neighborhood of
$\varphi_\text{end}\approx1.091$, and, as is typical of partial sums of asymptotic expansions,
get progressively worse, despite the fact that the prefactor in eq.~(\ref{eq:hhatfs}) enforces $h(0) = 0$.
As we mentioned in the Introduction, faced with an analogous problem (although with fewer terms) for the quadratic potential,
Liddle, Parsons and~Barrow~\cite{LID94} used [1/1] rational, multivariable Canterbury approximants.
However, and in light of the alternating sign apparent in the first terms
of eq.~(\ref{eq:hhatfs}) (for which we do not have a proof), we use
Pad\'e $[n_\text{SR}/n_\text{SR}+1]$ approximants to sum the formal series (a brief review of how to
compute these approximants is given in Appendix~\ref{ap:b}).
At the scale of figure~\ref{fig:srps}, already the $[2/3]$ Pad\'e approximant (not shown) would be superimposed
to the numerical integration (green curve). The errors of the Pad\'e approximants at 
$\varphi_\text{end}\approx1.091$ (black dot) decrease from $2.14\%$ for $n_\text{SR}=2$, to
$0.49\%$ for $n_\text{SR}=4$, and stabilize at $0.023\%$ for $n_\text{SR}=12$ and higher approximants,
which accounts for the difference between the numerical $h(\varphi_\text{end})$ of
our initial value problem and the value of the separatrix at that point 
to which the asymptotic series $h_\text{SR}(\varphi_\text{end})$ is being summed by the Pad\'e approximants.

\begin{figure}[tbp]
	\centering
        \includegraphics[width=7cm]{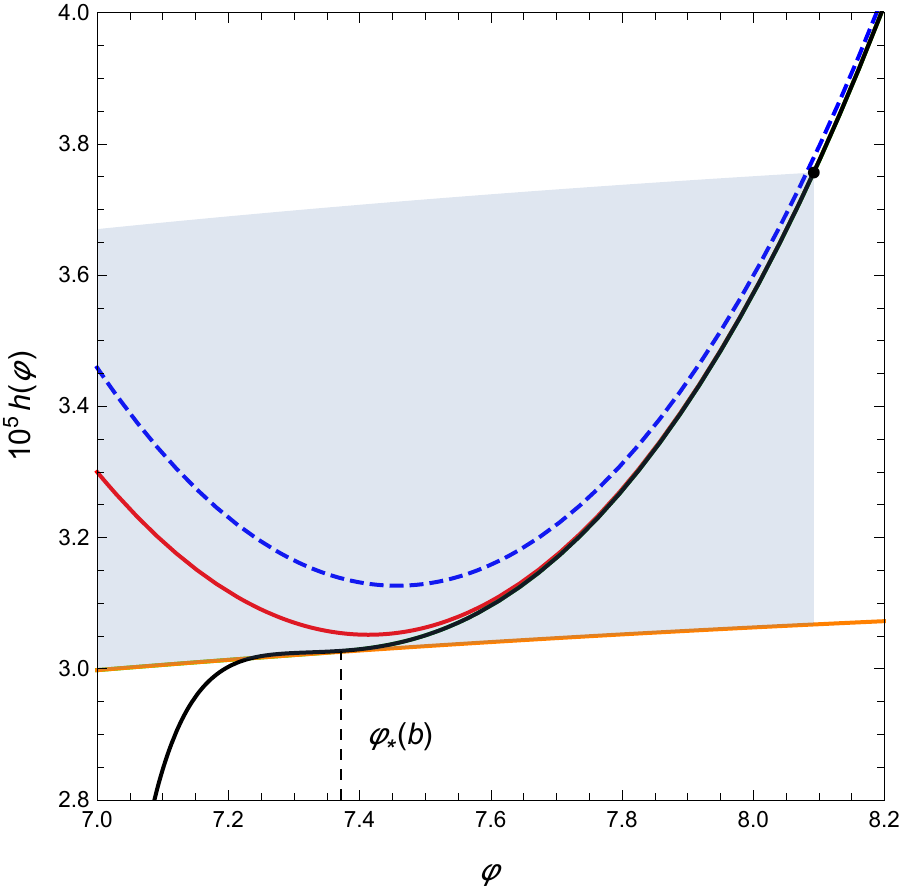}
    \caption{Numerical integration of $h(\varphi)$ (green curve), summation of $h_\text{SR}(\varphi)$
                  for $n_\text{SR}=12$,  (orange curve), summations of $h_\text{KD}(\varphi,b)$ for $n_\text{KD}=2$ (red curve) and
                  $n_\text{KD}=16$ (black curve), and first-order approximation from Ref.~\cite{CM19}
                  (blue, dashed curve). The vertical line at $\varphi_*(b) \approx 7.372$ marks the matching point.
                  The numerical (green) curve is barely visible because of the accurate approximations provided by
                  the SR (orange) curve up to $\varphi_*(b)$ and the KD (black) curve from $\varphi_*(b)$ onwards.
 	\label{fig:kd}}
\end{figure}

We consider next the KD asymptotic expansion. Figure~\ref{fig:kd} shows a neighborhood of $\varphi_\text{in}\approx 8.092$,
where the green curve (barely visible) is the same curve as in figure~\ref{fig:srps} and the orange curve is the Pad\'e-summed SR asymptotic
expansion with $n_\text{SR}=12$ (as we explained earlier, Pad\'e summation of $h_\text{SR}(\varphi)$ is unnecessary for these
values of $\varphi$ but essential in the neighborhood of $\varphi_\text{end}$), and the red and black curves are, respectively,
the $n_\text{KD}=2$ and $n_\text{KD}=16$ summations of the double series for $h_\text{KD}(\varphi,b)$ with $b=1.10817\times10^8$,
where, for consistency (and taking into account typical values of $\lambda$), the $n_\text{KD}$ summation is defined by
\begin{equation}
	\label{kd1pade}
	\hat{h}_\text{KD}^{[n_\text{KD}]}(y,b) = \frac{1}{b y^{\frac{1}{2\lambda}}}
	                 + \sum_{n=1}^{\lfloor 2 \lambda n_\text{KD} \rfloor} A^n b^{2n-1} \zeta_{k_n,k_n+1}(y) y^\frac{2n-1}{2\lambda},
\end{equation}
where $\lfloor x \rfloor$ denotes the greatest integer less than or equal to $x$,
\begin{equation}
	k_n = \lfloor n_\text{KD} - n/(2 \lambda)\rfloor,
\end{equation}
and $\zeta_{k_n,k_n+1}(y)$ denotes the $[k_n/k_n+1]$ Pad\'e approximant to $\zeta_n(y)$. The value of $b$
has been chosen so that the summed $h_\text{KD}(\varphi_0,b)=h(\varphi_0)=h_0$,
and the figure shows that the $n_\text{KD}=2$ approximation
(red curve) does not yet match the SR approximation (orange curve), but the $n_\text{KD}=16$ (black curve) does match it, and allows
us to solve for $\varphi_*(b)$ as the first local minimum of $(h_\text{KD}(\varphi,b) - h_\text{SR}(\varphi))^2$,
which turns out to be $\varphi_*(b) \approx 7.372$. Finally, the blue, dashed line represents a noteworthy first order
approximation derived by Chowdhury, Martin, Ringeval and Vennin in Ref.~\cite{CM19}, which in our
notation reads,
\begin{equation}
	\label{eq:cmrv}
	h(\varphi) \approx \frac{1}{2} ( c_+ e^{\varphi-\varphi_0} + c_- e^{-(\varphi-\varphi_0)} ),
\end{equation}
where
\begin{equation}
	c_\pm = h(\varphi_0) \pm \sqrt{h(\varphi_0)^2-v(\varphi_0)}.
\end{equation}
Figure~\ref{fig:kd} shows that this approximation is remarkably accurate for a first-order approximation,
but is not accurate enough to be matched with the SR approximation.

\begin{figure}[tbp]
	\centering
        \includegraphics[width=7cm]{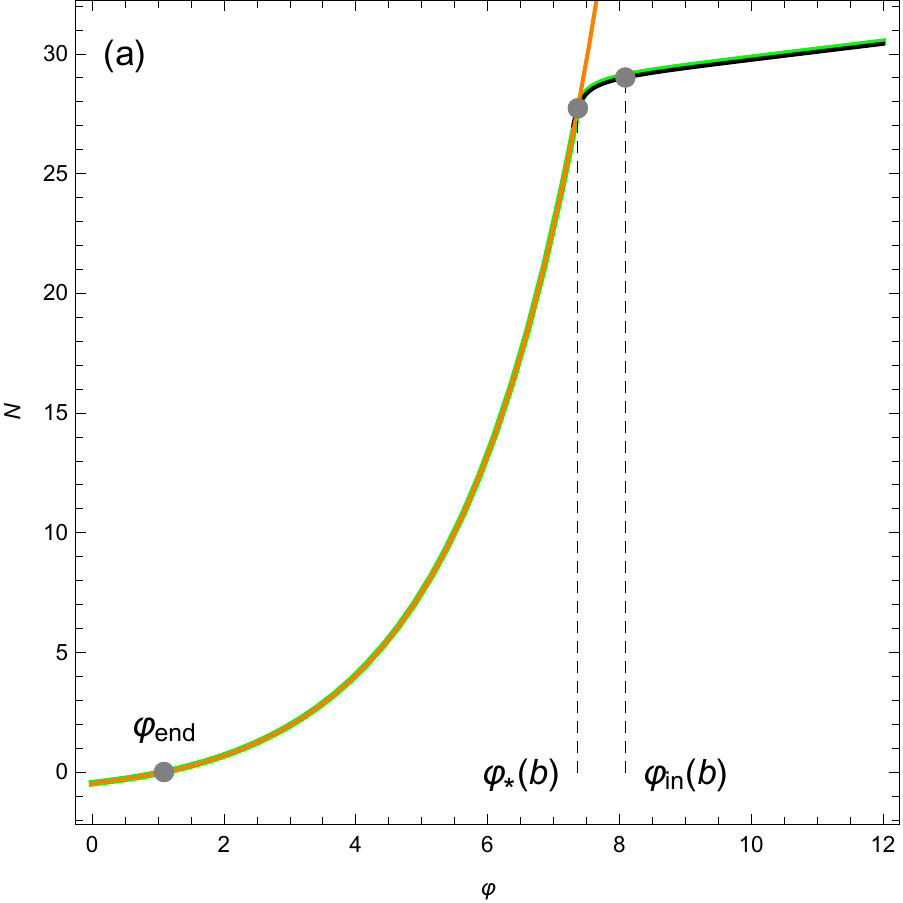}
        \hfill
        \includegraphics[width=7cm]{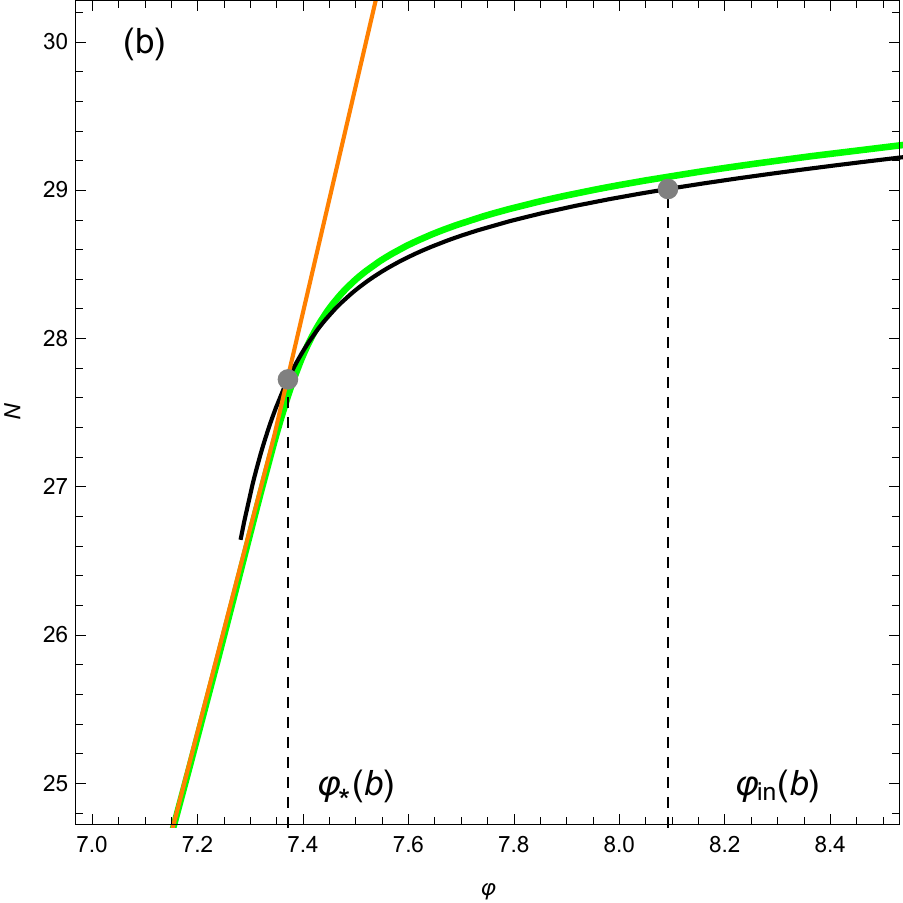}
    \caption{Numerical integration for $N(\varphi)$ (green curve), summation of $N_\text{SR}(\varphi,b)$ for $n_\text{SR}=16$,
                  (orange curve), and summation of $N_\text{KD}(\varphi,b)$ for $n_\text{KD}=16$ (black curve):
                  (a) on the inflation interval $[\varphi_\text{end},\varphi_\text{in}(b)]$;
                  (b) magnification of a neighborhood of $\varphi_*(b)$ showing how the crossing between the summations
                   mimic the inflection point of the numerical integration.
	\label{fig:a}}
\end{figure}

As we discussed above, the constant $\hat{N}_{\text{SR},0}$ is a choice of origin determined by eq.~(\ref{eqastarb}),
while matching the asymptotic expansions for the reduced Hubble parameter determines the constant
$\hat{N}_{\text{KD},0}$ in eq.~(\ref{akd1}) via eq.~(\ref{astarb}).
In practice we use an expression analogous to eq.~(\ref{kd1pade}),
\begin{equation}
	\label{akd1pade}
	\hat{N}_\text{KD}^{[n_\text{KD}]}(y,b) - \hat{N}_{\text{KD},0}(b)
	=
	- \frac{\log y}{6\lambda} - \sum_{n=1}^{\lfloor 2 \lambda n_\text{KD} \rfloor}A^n b^{2n}\xi_{k_n,k_n+1}(y) y^\frac{n}{\lambda},
\end{equation}
but at the risk of being repetitive we stress that the procedure to determine $\hat{N}_{\text{KD},0}$ is not a
matching---the expansions that are matched are the expansions for the reduced Hubble parameter.
To illustrate this point, in figure~\ref{fig:a} we show the result
of the numerical integration for $N(\varphi)$ (the green curve corresponding to the green curves in figures~\ref{fig:srps}
and~\ref{fig:kd}), the $n_\text{SR}=16$ summation of $N_\text{SR}(\varphi)$ (the orange curve), and the $n_\text{KD}=16$
summation of $N_\text{KD}(\varphi,b)$ (the black curve). The magnification in figure~\ref{fig:a}(b) shows how the matching at $\varphi_*(b)$
shown in figure~\ref{fig:kd} induces a crossing between the summations of the expansions for the number of e-folds that
mimics the inflection point of the numerical integration. Because of eq.~(\ref{hj3}), the numerical integration for $N(\varphi)$
is more sensitive to the initial conditions than the numerical integration for $h(\varphi)$, but the Pad\'e-summed asymptotic
expansion for $N_\text{KD}(\varphi,b)$ gives a remarkably accuracy up to the crossing point.

\section{Applications\label{sec:app}}
\subsection{The total amount of inflation as a function of the initial condition}
The approximations to the reduced Hubble parameter $h(\varphi)$ and to the number of e-folds $N(\varphi)$
on the whole inflation interval $[\varphi_\text{end},\varphi_\text{in}(b)]$ that the matching procedure discussed
in the previous section produces allows us to derive ensuing approximations to several relevant
magnitudes as functions of the parameter $b$ (or, equivalently, of the initial conditions), e.g., the values
of the inflaton for which a solution enters and exits the inflation region or the total amount of inflation, or,
conversely, to find the initial conditions for a solution to correspond to a previously fixed total amount of inflation.
 
\begin{figure}[tbp]
	\centering
        \includegraphics[width=7.5cm]{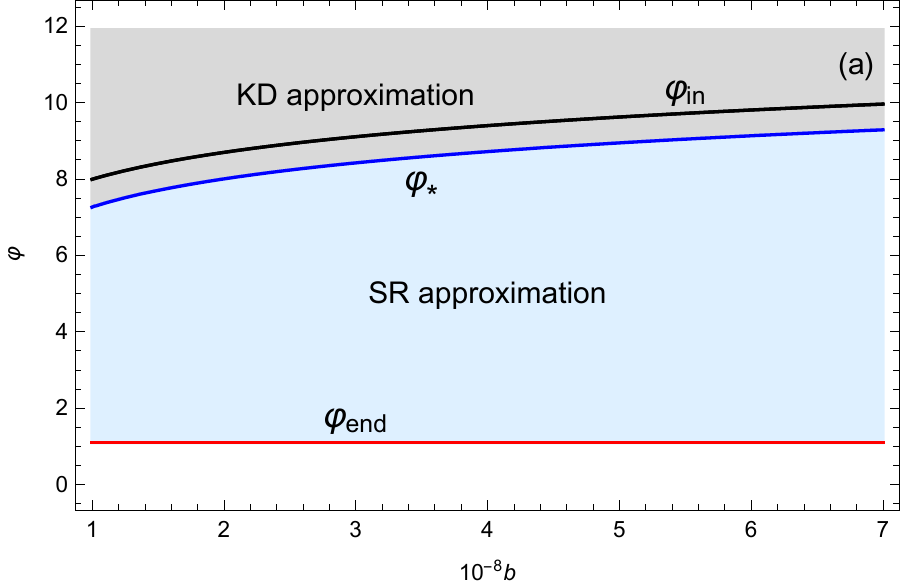}
        \hfill
        \includegraphics[width=7.5cm]{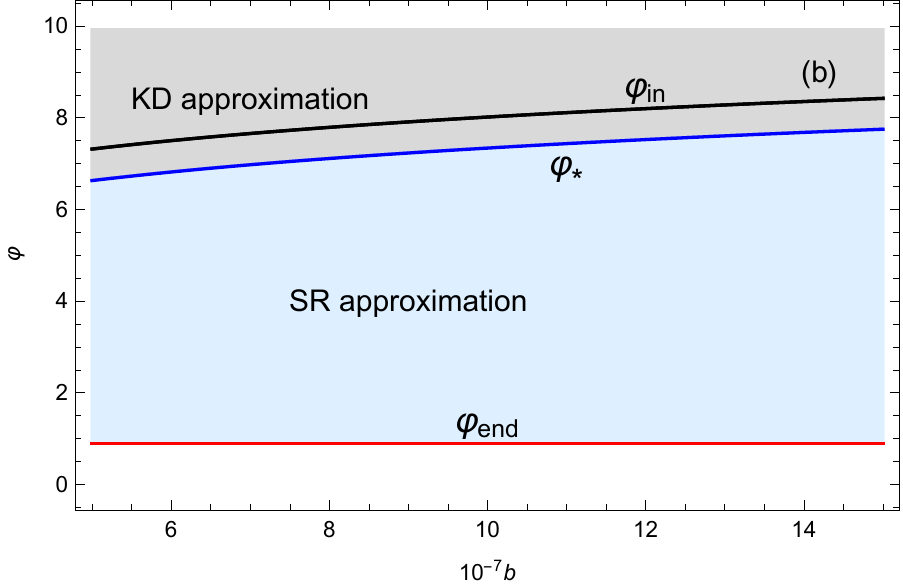}\\ 
        \vspace{0.5cm}
        \includegraphics[width=7.5cm]{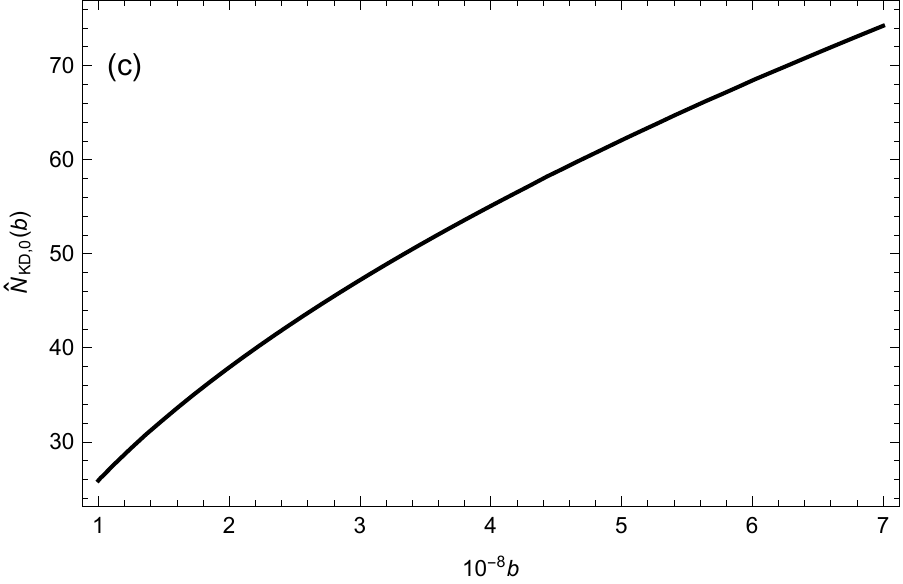}
        \hfill
        \includegraphics[width=7.5cm]{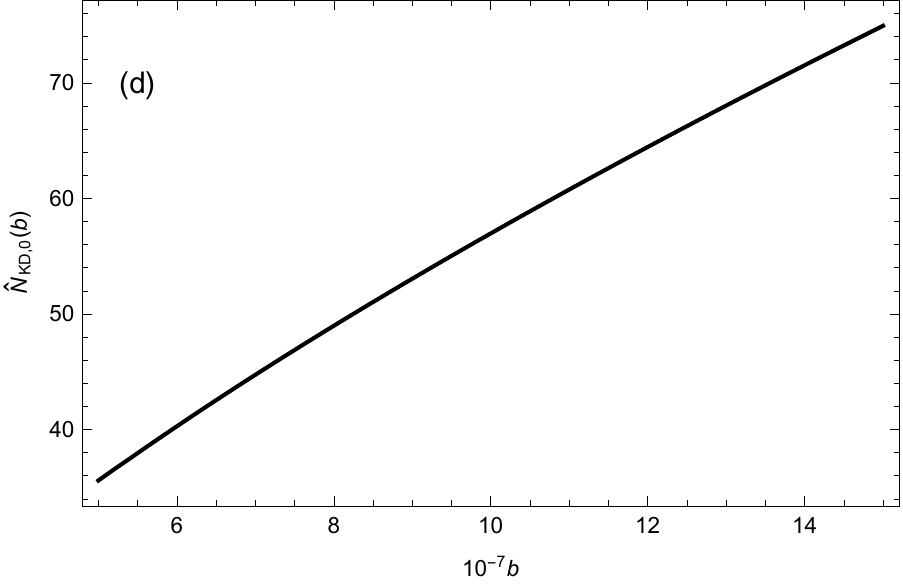}
    \caption{Results of the matching of the $n_\text{SR}=16$ and $n_\text{KD}=16$ approximants  for the T-models
                   of Table~\ref{tab:param} as functions of the parameter $b$:
                  (a), (b) $\varphi_\text{end}$ (constant, red line), $\varphi_*(b)$ (blue line) and $\varphi_\text{in}(b)$ (black line)
                  for Example~I and Example~II respectively;
                  (c), (d) $\hat{N}_{\text{KD},0}(b)$ for Example~I and Example~II respectively.
    \label{fig:regions}}
\end{figure}

\begin{figure}[tbp]
	\centering
        \includegraphics[width=7.5cm]{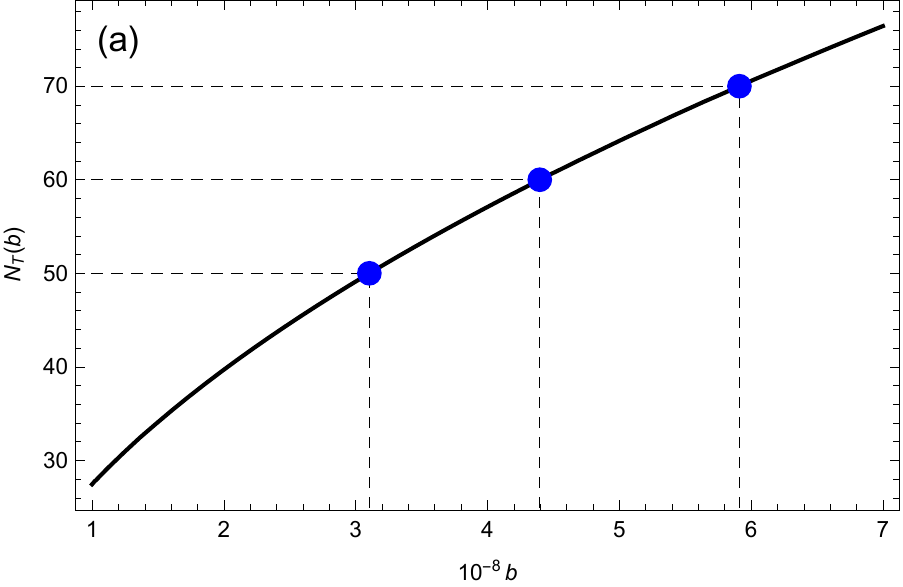}
        \hfill
        \includegraphics[width=7.5cm]{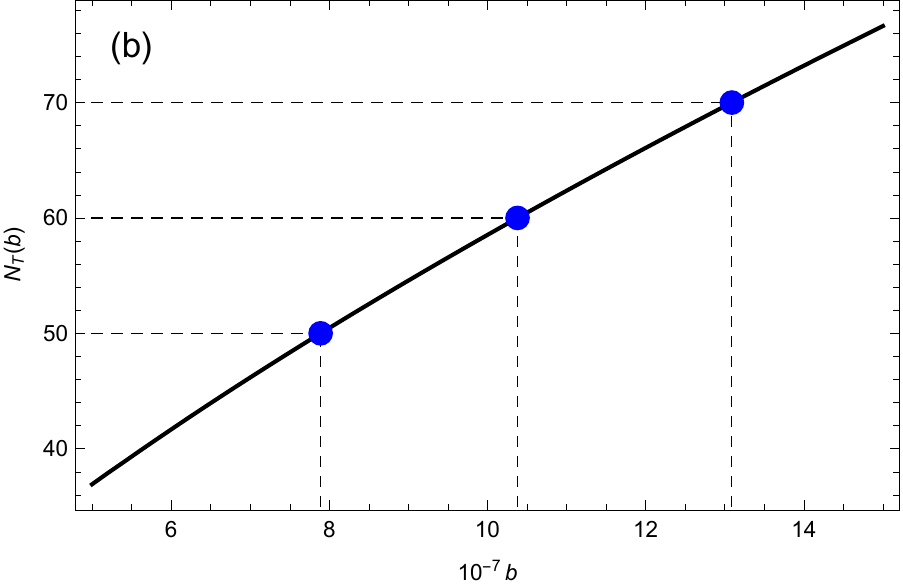}
    \caption{Approximate total amount of inflation $N_T(b)$ as a function of the parameter $b$
                  for $n_\text{SR} = 16$ and $n_\text{KD} = 16$.
                  The dots mark $N_T = 50$, $60$ and $70$:
                  (a) Example~I in Table~\ref{tab:param}; (b) Example~II in Table~\ref{tab:param}.
                  \label{fig:nt}}
\end{figure}

\begin{table}[tbp]
\centering
\begin{tabular}{cccc}
\hline
  Example & $A$ & $m$ & $\lambda$ \\
\hline
     I & $10^{-9}$ & $1$           & $1/\sqrt{15}$  \\
     II & $10^{-9}$ & $7/8$        & $1/3$            \\
\hline    
\end{tabular}
\caption{\label{tab:param}
              Parameters for the T-models used in section~\ref{sec:app}.
              The values of $m$ and $\lambda$ for Example~I and~II are taken from Refs.~\cite{AK18}, and~\cite{CA15},
              respectively. The values of $A$ are derived from typical values of $\Lambda$.}
\end{table}
We illustrate these applications with two examples whose parameters, summarized in Table~\ref{tab:param},
are chosen according to the following considerations. Our first example has $m=1$ and  $\alpha=5/3$
(or, using eq.~(\ref{redal}), $\lambda=1/\sqrt{15}$)~\cite{AK18}.
The corresponding values of $\Lambda$ found in the literature range from
$\Lambda=\alpha 10^{-10}\mathrm{M}_\mathrm{Pl}^2$~\cite{CA15,AK18} to
$\Lambda=10^{-8}\mathrm{M}_\mathrm{Pl}^2$~\cite{SD20}. Using again eq.~(\ref{redal}), we have taken as a typical
value $A=10^{-9}$. This is the potential shown in figure~\ref{fig:v}, with the corresponding phase portrait
shown in figures~\ref{fig:pp} and~\ref{fig:newfig}, and used for illustrating the several steps of the matching
procedure in figures~\ref{fig:srps}--\ref{fig:a}.
Our second example, taken from Ref.~\cite{CA15}, has the noninteger value of $m=7/8$,
$\alpha=1$ (i.e., $\lambda=1/3$), and again the typical value $A=10^{-9}$.
Before proceeding to the applications proper, in Fig~\ref{fig:regions} we summarize the results of the matching
for both examples as a function of the parameter $b$ for $n_\text{SR}=n_\text{KD}=16$.
The constants $\hat{N}_{\text{SR},0}$ (determined via eq.~(\ref{eqastarb}) for Examples~I and~II are
$-1.69$ and $-1.31$, respectively. Figures~\ref{fig:regions}(a) and~(b) show the values of $\varphi_\text{end}$ (constant),
$\varphi_*(b)$ and  $\varphi_\text{in}(b)$ for Example~I and Example~II respectively. We recall that the Pad\'e
summation provides an SR approximation to the separatrix accurate on $[\varphi_\text{end},\infty)$, and
a KD approximation for a particular, $b$-dependent solution valid on
$[\varphi_*(b),\infty)$. These intervals are illustrated by the shaded regions in the figure.
The values for $\hat{N}_{\text{KD},0}(b)$ defined in eq.~(\ref{akd1}) and computed according to eq.~(\ref{astarb})
for Examples~I and~II are shown in figures~\ref{fig:regions}(c) and~(d) respectively.
The total amount of inflation or number of e-folds during the inflation period is
\begin{equation}
	\label{Na}
	N_T = \log\frac{a(t_\text{end})}{a(t_\text{in})} = N(\varphi_\text{in})-N(\varphi_\text{end}) = N(\varphi_\text{in}),
\end{equation}
which we approximate by
\begin{equation}
	N_T \approx \hat{N}_\text{KD}^{[n_\text{KD}]}(e^{-2\lambda\varphi_\text{in}(b)}),
\end{equation}
and should be close to $N_T\approx 60$ \cite{BA09,BA12,DO03,MA18}.  Figure~\ref{fig:nt} shows the result of a calculation
with $n_\text{SR} = 16$ and $n_\text{KD} = 16$ for the two examples in Table~\ref{tab:param}, and Table~\ref{tab:nt506070}
shows a comparison of the results obtained from the asymptotic expansions for $N_T(b)=50$, $60$ and $70$
(the marked points on figure~\ref{fig:nt}) with the results of a numerical integration.
More concretely, for each value of $N_T$ we find the corresponding value of $b$ in figure~\ref{fig:nt} and perform a numerical
integration of eqs.~(\ref{hj1}) and~(\ref{hj3}) with initial conditions at $\varphi_0=10$ given by
$h(\varphi_0) = h_\text{KD}(\varphi_0,b)$ and $N(\varphi_0) = N_\text{KD}(\varphi_0,b)$. In all the cases we have tested,
the errors are below $0.5\%$, i.e., well below one e-fold. Note also that the number of e-folds in the fast-roll stage
that the Pad\'e summation of the KD expansion allows us to reach is only of about 1.2,
which is consistent with the observation made by Goldwirth and Piran~\cite{GP92} after their eq.~(4.13).

\begin{table}[tbp]
\centering
\begin{tabular}{ccccccc}
\hline
     & \multicolumn{3}{c}{Example~I} & \multicolumn{3}{c}{Example~II} \\
     \cline{2-4} \cline{5-7}
   $N_T(b)$ & $b$ & $N_T^\text{(num)}(b)$ & $N_\text{SR}(\varphi_*(b))$ & $b$ & $N_T^\text{(num)}(b)$  & $N_\text{SR}(\varphi_*(b))$\\
\hline
50 & $3.10\times 10^8$ & 50.03  & 48.98 & $7.89\times 10^7$ & 50.03  & 48.83 \\
60 & $4.40\times 10^8$ & 60.00  & 58.92 & $1.04\times 10^8$ & 60.06  & 58.84 \\
70 & $5.91\times 10^8$ & 70.00 &  68.93 & $1.31\times 10^8$ & 70.09  & 68.85 \\
\hline
\end{tabular}
\caption{\label{tab:nt506070}
              Comparison between the total amount of inflation $N_T(b)$
              with $n_\text{SR}=16$ and $n_\text{KD}=16$ and the result of a numerical integration from the corresponding
              initial condition $b$. $N_\text{SR}(\varphi_*(b))=N_\text{KD}(\varphi_*(b),b)$ denotes the number of e-folds in the SR stage.}
\end{table}
\subsection{Second-order SR approximation to $n_s$ and $r$ for T-models}
The spectral index $n_s$ and the tensor-to-scalar ratio $r$ for T-models in the first-order SR approximation are given by
\begin{equation}
	\label{eq:ns0}
	n_s = 1 - \frac{2}{N},
\end{equation}
and
\begin{equation}
	\label{eq:r0}
	r = \frac{4}{3 \lambda^2 N^2},
\end{equation}
although the latter equation is often written in terms of $\alpha$ instead of
$\lambda$~\cite{KL13,KL13b,KALL15,GA15,CA15,AK18,AC21}.  Note in particular that the first-order approximation
to $n_s$ is independent of the parameters of the model. In this section, and taking advantage of our former SR results,
we compute the spectral index $n_s(N)$ and the tensor-to-scalar ratio $r(N)$ to second order in the SR approximation or, more precisely,
to order $1/N^2$ and $1/N^3$ respectively. (We remark that this computation neither uses nor depends in any way
of the KD series---the resulting expressions are pure SR results.)

To this aim, we consider the $O(\epsilon_i^3)$ expressions~\cite{LE02},
\begin{equation}
	\label{eq:ns}
	n_s = 1 - 2 \epsilon_1 - \epsilon_2 - 2 \epsilon_1^2 - (2C+3) \epsilon_1 \epsilon_2 - C \epsilon_2 \epsilon_3,
\end{equation}
and
\begin{equation}
	\label{eq:r}
	r = 16 \epsilon_1 \left[ 1 + C \epsilon_2 + \left( C - \frac{\pi^2}{2} + 5 \right) \epsilon_1 \epsilon_2
	                                       + \left( \frac{C^2}{2} - \frac{\pi^2}{8} + 1 \right) \epsilon_1^2
	                                       + \left( \frac{C^2}{2} - \frac{\pi^2}{24} \right) \epsilon_2 \epsilon_3 \right],
\end{equation}
where $C$ can be written in terms of the Euler-Mascheroni constant $\gamma$,
\begin{equation}
	C = \gamma + \log 2 - 2,
\end{equation}
and the slow-roll parameters $\epsilon_1$, $\epsilon_2$, and $\epsilon_3$ can be written as functions of $N'(\varphi)$
(cf.~eq.~(\ref{hj3})) and its derivatives,
\begin{equation}
	\label{eh1}
	\epsilon_1(\varphi) = \frac{1}{3} \frac{1}{N'(\varphi)^2},
\end{equation}
\begin{equation}
	\label{eh2}
	\epsilon_2(\varphi) = - 2 \frac{d}{d\varphi}\frac{1}{N'(\varphi)},
\end{equation}
\begin{equation}
	\label{eh23}
	\epsilon_2(\varphi)\epsilon_3(\varphi) = \frac{2}{N'(\varphi)} \frac{d^2}{d\varphi^2}\frac{1}{N'(\varphi)}.
\end{equation}
We will see later that although all the terms in eq.~(\ref{eq:ns}) are required, only the first two terms
in eq.~(\ref{eq:r}) need to be considered, i.e.,
\begin{equation}
	\label{eq:rtwoterms}
	r = 16 \epsilon_1 (1 + C \epsilon_2 ).
\end{equation}

Since all our expansions in this section pertain to the SR stage, hereafter we drop the subindex ``SR'' but 
keep the circumflex accent to denote magnitudes in the variable $y$. By substituting eq.~(\ref{eq:ahat})
and the second order approximation
\begin{equation}
	\label{eq:ehh0}
	\hat{N}(y) = \hat{N}_0+\frac{1}{24m\lambda^2}\left(1+\frac{1}{y}\right)-\frac{1}{3}\log y+\frac{16}{3}m\lambda^2y+O(y^2),
\end{equation}
into eqs.~(\ref{eh1})--(\ref{eh23}) we find that,
\begin{equation}
	\label{eq:ehh1}
	\hat{\epsilon}_1(y) = \frac{48 m^2 \lambda^2 y^2}{(1-y^2)^2}
	                                                   ( 1 - 16 m \lambda^2 y + O(y^2)),
\end{equation}
\begin{equation}
	\label{eq:ehh2}
	\hat{\epsilon}_2(y) = \frac{48 m \lambda^2 y (1+y^2)}{(1-y^2)^2}
	                                                  ( 1 - 16 m \lambda^2 y + O(y^2)),
\end{equation}
\begin{equation}
	\label{eq:ehh23}
	\hat{\epsilon}_2(y) \hat{\epsilon}_3(y)
	= 
	\frac{1152 m^2 \lambda^4 y^2 (1+6y^2+y^4)}{(1-y^2)^4}
	(1 + O(y)).
\end{equation}
Note that the second order approximation~(\ref{eq:ehh0}) differs from the standard SR
approximation~(\ref{eq:nnyy}) in both the logarithmic term and the first term in the series in eq.~(\ref{eq:aSR}),
and that the prefactors in eqs.~(\ref{eq:ehh1})--(\ref{eq:ehh23}) are precisely the first-order SR results.

Then, we need to find the value of $y_\text{end}$ corresponding to the end of the inflationary stage to second order in the SR
approximation, i.e., to solve
\begin{equation}
	\label{eq:yend1}
	\hat{\epsilon}_1(y_\text{end}) = 1
\end{equation}
to second order. The first order approximation is the suitable solution of the simple biquadratic equation
\begin{equation}
	\frac{48 m^2 \lambda^2 y^2}{(1-y^2)^2} = 1,
\end{equation}
namely
\begin{equation}
	\label{eq:yendSR1}
	y_{\text{end},1} = \sqrt{1+12 m^2\lambda^2} - 2\sqrt{3} m \lambda,
\end{equation}
and the second order correction is obtained by substituting $y = y_{\text{end},1} + y_{\text{end},2}$ into
eq.~(\ref{eq:yend1}) and expanding to first order in $y_{\text{end},2}$ using eq.~(\ref{eq:ehh1}). The full result turns out to be,
\begin{equation}
	\label{eq:yendsol}
	y_\text{end} = (\sqrt{1+12 m^2\lambda^2} - 2\sqrt{3} m \lambda)
	                       \left(
	                       		1 + \frac{48 m^2 \lambda^3}{\sqrt{1+12 m^2\lambda^2}(6 m \lambda + \sqrt{3} \sqrt{1+12 m^2\lambda^2})}
	                       \right).
\end{equation}
Next we need to expand the solution of the equation
\begin{equation}
	\label{eq:nysec}
	\hat{N} - \hat{N}_0= \frac{1}{24 m \lambda^2}\left(\frac{1}{y}+y\right) - \frac{1}{3} \log y+\frac{16}{3}m\lambda^2y,
\end{equation}
for $y$ as a function of $\hat{N}$ as $\hat{N}\rightarrow\infty$ to order $1/\hat{N}^2$. In order to do this we consider first
the two leading terms in the right-hand side of eq.~(\ref{eq:nysec}),
\begin{equation}
	\label{eq:ny2nd}
	\hat{N} - \hat{N}_0= \frac{1}{24 m \lambda^2 y} - \frac{1}{3} \log y,
\end{equation}
where
\begin{equation}
	\label{eq:n0yend}
	\hat{N}_0 =  - \frac{1}{24 m \lambda^2 }\left(\frac{1}{y_\text{end}}+y_\text{end}\right) + \frac{1}{3} \log (y_\text{end})
	 -\frac{16}{3}m\lambda^2y_\text{end},
\end{equation}
so that $\hat{N}(y_\text{end}) = 0$. Equation~(\ref{eq:ny2nd}) can be solved exactly in terms of
Lambert's $W$ function~\cite{OL10},
\begin{equation}
	y = \frac{1}{8 m \lambda^2 W (e^{3(\hat{N} - \hat{N}_0)}/8 m \lambda^2)},
\end{equation}
which leads to the expansion (cf.~again Ref.~\cite{OL10}),
\begin{equation}
	\label{eq:yN2}
	y = \frac{1}{24 m \lambda^2} \left[ \frac{1}{\hat{N}} + \frac{\log \hat{N}}{3 \hat{N}^2} + \frac{3 \hat{N}_0 + \log (24m\lambda^2)}{3 \hat{N}^2}
	+ O\left( \frac{(\log \hat{N})^2}{\hat{N}^3}\right) \right].
\end{equation}
To check if this equation is in fact the solution of eq.~(\ref{eq:nysec}) to the order stated we substitute an expansion of the form
\begin{equation}
	\label{eq:newexp}
	y = \frac{1}{24 m \lambda^2} \left[ \frac{1}{\hat{N}}
	    + \frac{\log \hat{N}}{3 \hat{N}^2} + \frac{q}{\hat{N}^2}
	    + \frac{p_2(\log \hat{N})^2}{\hat{N}^3}+ \frac{p_1\log \hat{N}}{\hat{N}^3}
	    + \frac{p_0}{\hat{N}^3}+ O\left( \frac{(\log \hat{N})^3}{\hat{N}^4}\right) \right]
\end{equation}
into eq.~(\ref{eq:nysec}), equate to zero in the resulting equation the coefficients of $\hat{N}^0$
and $\hat{N}^{-1}(\log \hat{N})^k$, $k=0,1,2$ and find that
\begin{equation}
	q=\hat{N}_0 + \frac{1}{3}\log (24m\lambda^2),
\end{equation}
\begin{equation}
	p_0=q^2-\frac{q}{3}+\frac{2}{9}+\frac{1}{576m^2\lambda^4},
\end{equation}
\begin{equation}
	p_1 = \frac{2}{3}q-\frac{1}{9},
\end{equation}
\begin{equation}
	p_2 = \frac{1}{9}.
\end{equation}
Thus, we confirm that the solution of eq.~(\ref{eq:nysec}) to order $1/\hat{N}^2$ is indeed
given by eq.~(\ref{eq:yN2}). Incidentally, we mention that the first term in which the expansions
of the solutions of eqs.~(\ref{eq:nysec}) and~(\ref{eq:ny2nd}) differ is the term in $1/\hat{N^3}$. 

Finally, by substituting eq.~(\ref{eq:yN2}) into eqs.~(\ref{eq:ehh1})--(\ref{eq:ehh23}), these in turn into eqs.~(\ref{eq:ns}) and~(\ref{eq:rtwoterms}),
and dropping the circumflex accent of $N$, which is now not considered a function but the variable, we arrive at,
\begin{equation}
	\label{eq:rebmns}
	n_s = 1
 		 - \frac{2}{{N}}
	          - \frac{2}{3} \frac{\log{N}}{{N}^2}
	          - \frac{2}{{N}^2} \left( \frac{1}{12 \lambda^2} + \hat{N}_0 + \frac{\log (24m\lambda^2) - 2}{3} + C \right)
	          + O\left(\frac{(\log{N})^2}{{N}^3}\right),
\end{equation}
and
\begin{equation}
	\label{eq:rebmr}
	r = \frac{4}{3\lambda^2{N}^2}
	     + \frac{8}{9\lambda^2} \frac{\log{N}}{{N}^3}
	     + \frac{8}{9\lambda^2{N}^3} (3\hat{N}_0 + \log (24m\lambda^2) -1 + 3 C)
	     + O\left(\frac{(\log{N})^2}{{N}^4}\right).
\end{equation}
Note the presence of logarithmic terms in the expansions that we mentioned
in the Introduction and that are missed in Refs.~\cite{KA13,OO16,GE21}, and the fact that the first correction
to the $n_s$ in eq.~(\ref{eq:ns0}) is still independent of the parameters of the T-model, which appear only in the next term.
At the risk of being repetitive, we remark that eqs.~(\ref{eq:rebmns}) and~(\ref{eq:rebmr}) follow from the standard SR
expansion~(\ref{eq:SSREX})---our method of computing this expansion by adapting it to the analytic
form of the potential~(\ref{Tmodel}) makes their derivation both practical and systematic.
\subsection{Accuracy of the first- and second-order SR approximations}
In this section we present some numerical results that illustrate the accuracy  of the first- and second-order (purely)
SR approximations to the spectral index and tensor to scalar ratio as functions of the number of e-folds and of the parameters of the model.
All these results pertain to the T-model given in eq.~(\ref{redv}) with $A = 10^{-9}$.

\begin{figure}[tbp]
	\centering
        \includegraphics[width=7.4cm]{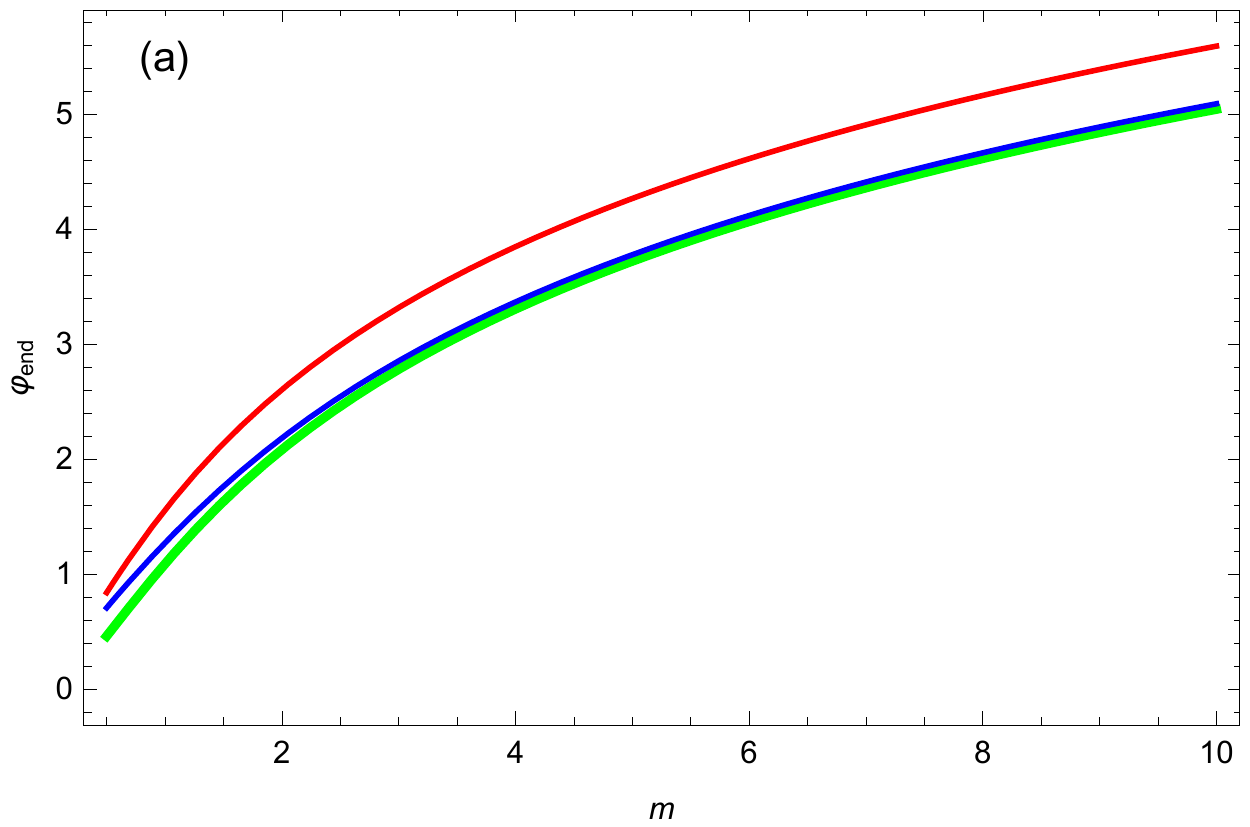}
        \hfill
        \includegraphics[width=7.5cm]{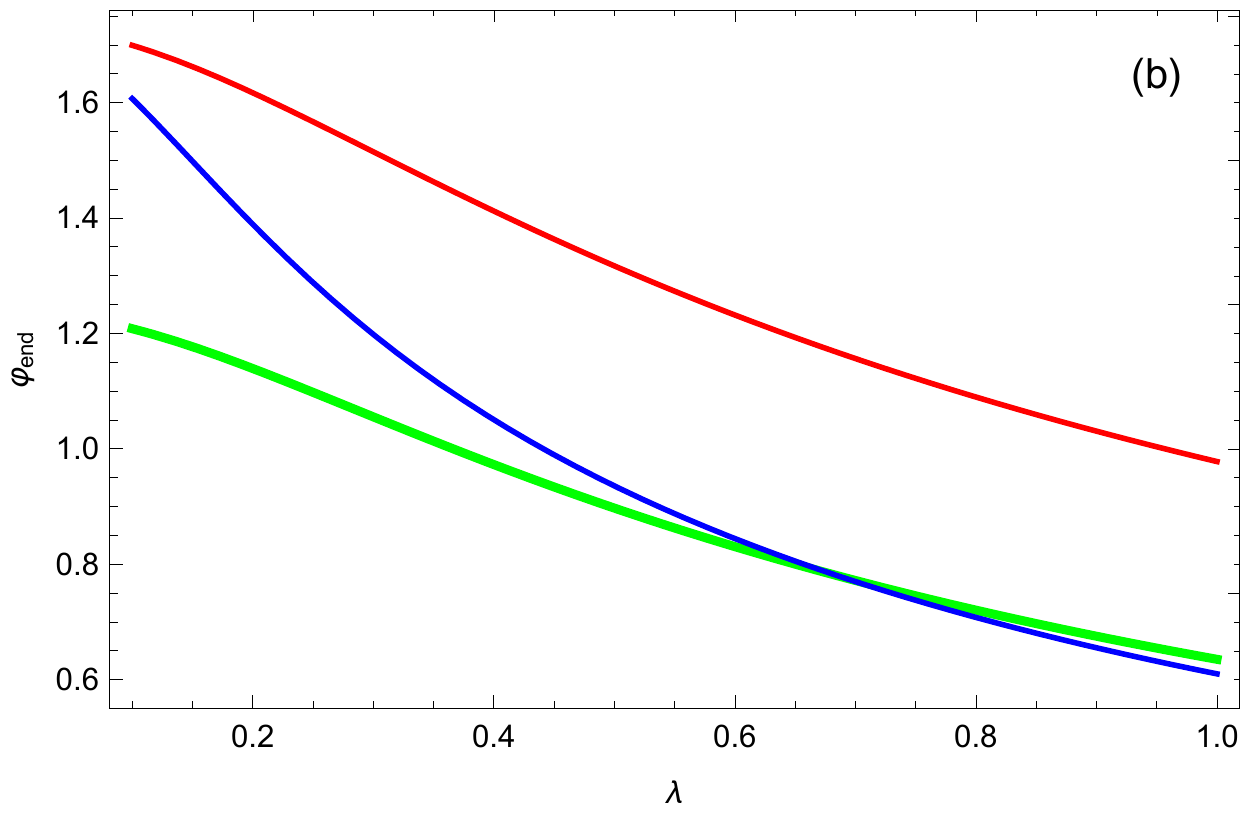}
    \caption{End of inflation $\varphi_\text{end}$ for T-models with $A = 10^{-9}$ and initial condition
                  $h(\varphi_0)=4.4\times 10^{-5}$ at $\varphi_0=10$. The red curves are the results given by
                  the first-order SR approximation eq.~(\ref{eq:yendSR1}), the blue curves are the results
                  given by the second-order SR approximation eq.~(\ref{eq:yendsol}), and the green
                  curves are the results of numerical integrations:
                  (a) as a function of the parameter $m$ for a fixed value $\lambda = 1/\sqrt{15}$;
                  (b) as a function of the parameter $\lambda$ for a fixed value of $m=1$.
	\label{fig:fendmlambda}}
\end{figure}

The first magnitude of interest is the value of the inflaton at the end of inflation $\varphi_\text{end}$, because its
value enters the derivation via eqs.~(\ref{eq:nysec}) and~(\ref{eq:n0yend}). In  figure~\ref{fig:fendmlambda} we compare the values of
$\varphi_\text{end}$ given by the first-order SR approximation eq.~(\ref{eq:yendSR1}) (red curve),
by the second-order SR approximation eq.~(\ref{eq:yendsol}) (blue curve), and by a numerical integration (green curve).
Figure~\ref{fig:fendmlambda}~(a) shows $\varphi_\text{end}$ as a function of the parameter
$m$ for a fixed value $\lambda = 1/\sqrt{15}$, and figure~\ref{fig:fendmlambda}~(b) shows $\varphi_\text{end}$
as a function of the parameter $\lambda$ for a fixed value $m=1$. These ranges of $m$ and $\lambda$ cover
the cases of physical interest, and across all of them the second-order SR approximation is clearly
more accurate than the first-order SR approximation.

\begin{figure}[tbp]
	\centering
        \includegraphics[width=7.3cm]{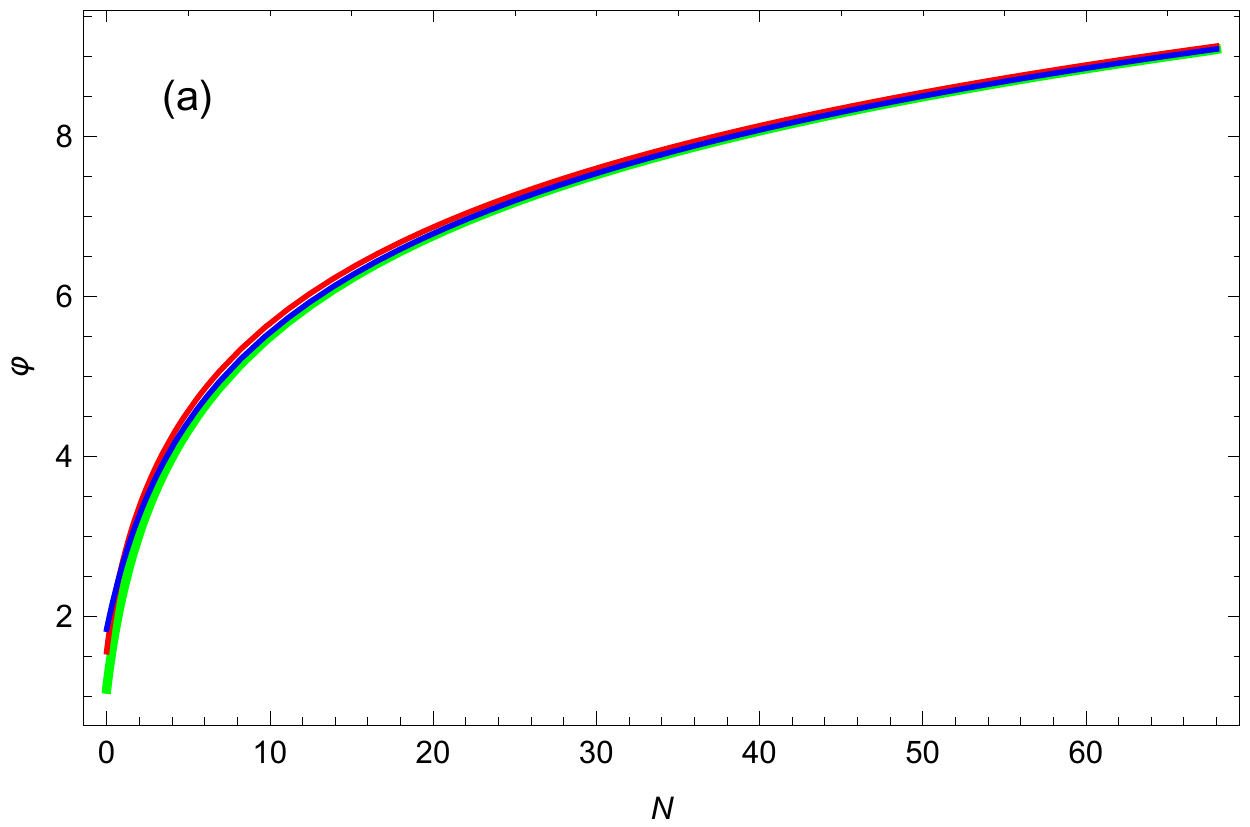}
        \hfill
        \includegraphics[width=7.6cm]{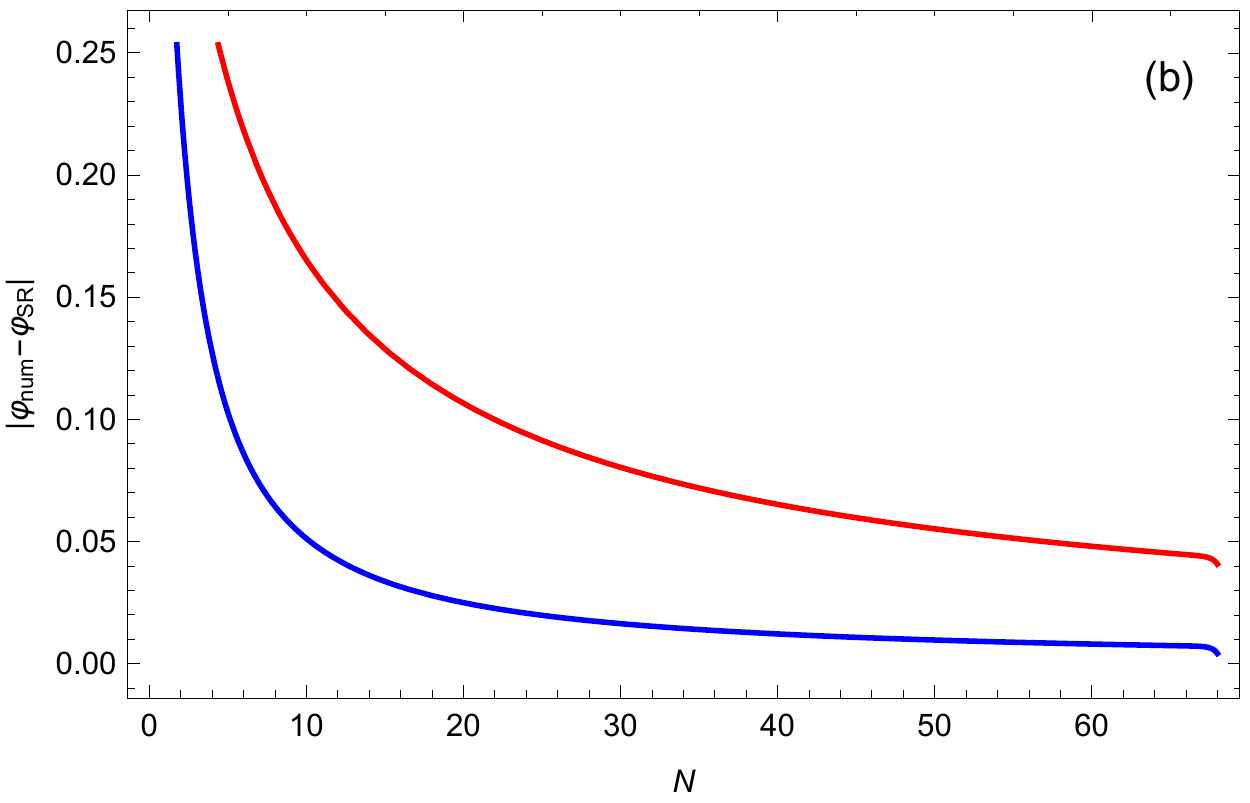}
    \caption{ (a) Inflaton $\varphi$ as a function of the number of e-folds $N$ for a T-model with $A = 10^{-9}$,
                        $m=1$, $\lambda = 1/\sqrt{15}$ and initial condition $h(\varphi_0)=4.4\times 10^{-5}$ for $\varphi_0=10$ as a function
                        of the number of e-folds $N$. The red curve is the first-order SR approximation, the blue curve is the second-order
                        SR approximation, and the green curve is the result of a numerical integration.
                   (b) Absolute values of the differences between the numerical value of the inflaton and its first order (red curve)
                         and second order (blue curve) SR approximations.
          \label{fig:fn}}
\end{figure}

In figure~\ref{fig:fn}~(a) we show the value of the inflaton $\varphi$ as a function of the number of e-folds $N$ for
$m=1$ and $\lambda = 1/\sqrt{15}$. The red curve is the first-order SR approximation, the blue curve is the second-order
SR approximation, and the green curve is the result of a numerical integration. The order of magnitude of $\varphi$ in
this figure does not allow to appreciate the respective accuracies. Therefore, in figure~\ref{fig:fn}~(b) we show for the
corresponding differences $|\varphi_\mathrm{num}-\varphi_\mathrm{SR,1}|$ (red curve) and
$|\varphi_\mathrm{num}-\varphi_\mathrm{SR,2}|$ (blue curve). The second-order SR approximation is clearly more
accurate.

\begin{figure}[tbp]
	\centering
        \includegraphics[width=7.4cm]{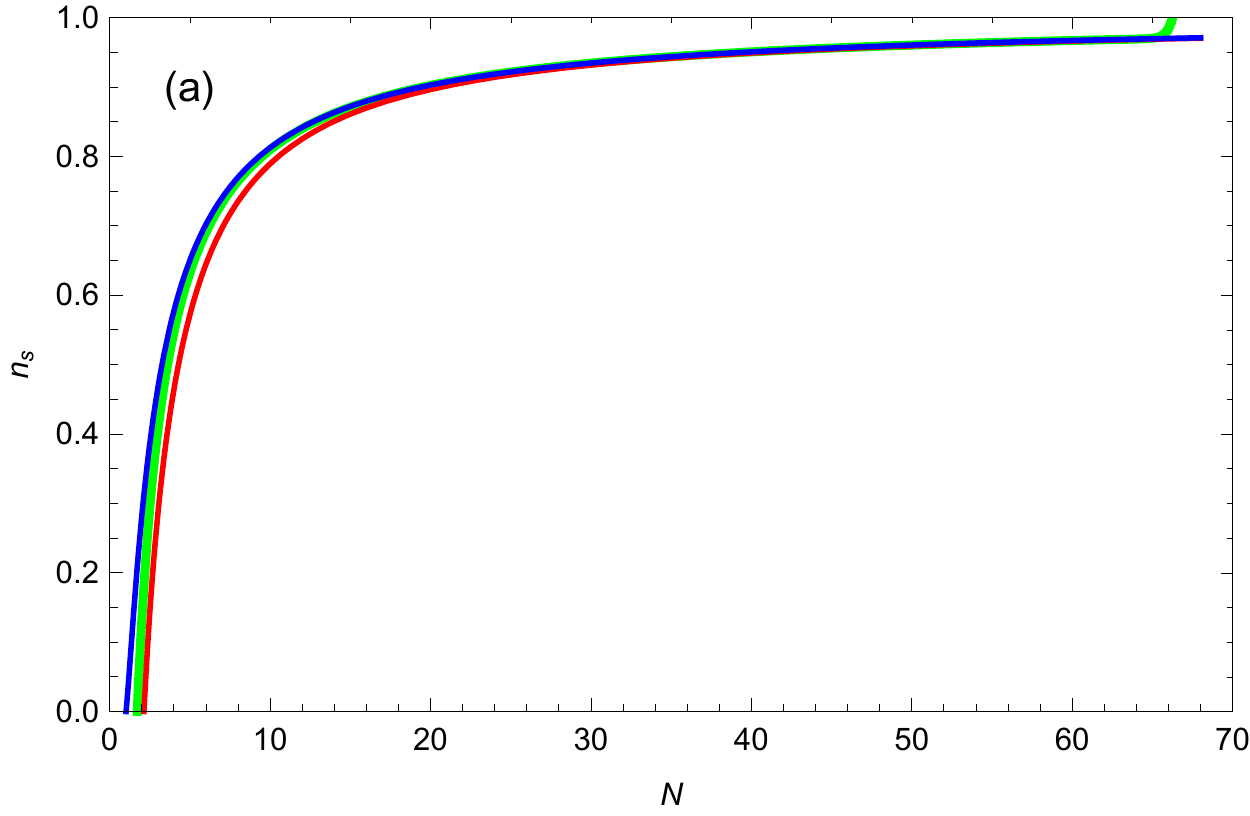}
        \hfill
        \includegraphics[width=7.6cm]{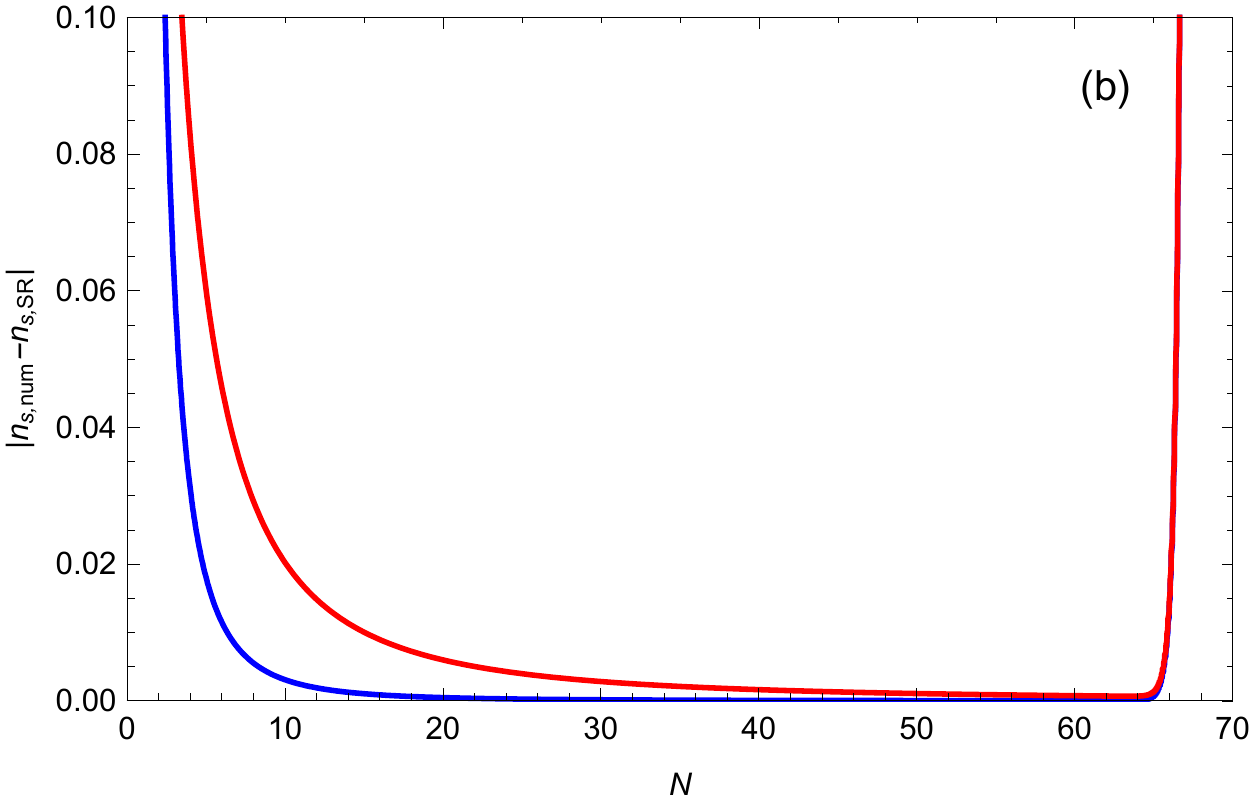}
    \caption{ (a) Spectral index $n_s$ for a  T-model with $A = 10^{-9}$, $m=1$, $\lambda = 1/\sqrt{15}$ and initial condition
                        $h(\varphi_0)=4.4\times 10^{-5}$ for $\varphi_0=10$ as a function of the number of e-folds $N$. The red curve is the
                        result given by the first-order SR approximation, the blue curve is the result
                        given by the second-order SR approximation, and the green curve is the
                        result of a numerical integration.
                   (b) Absolute values of the differences between the numerical value of the spectral index and its first order (red curve)
                         and second order (blue curve) SR approximations. In the region beyond $N=65$ both
                         curves are superimposed.                   
	\label{fig:ns}}
\end{figure}

\begin{figure}[tbp]
	\centering
        \includegraphics[width=7.35cm]{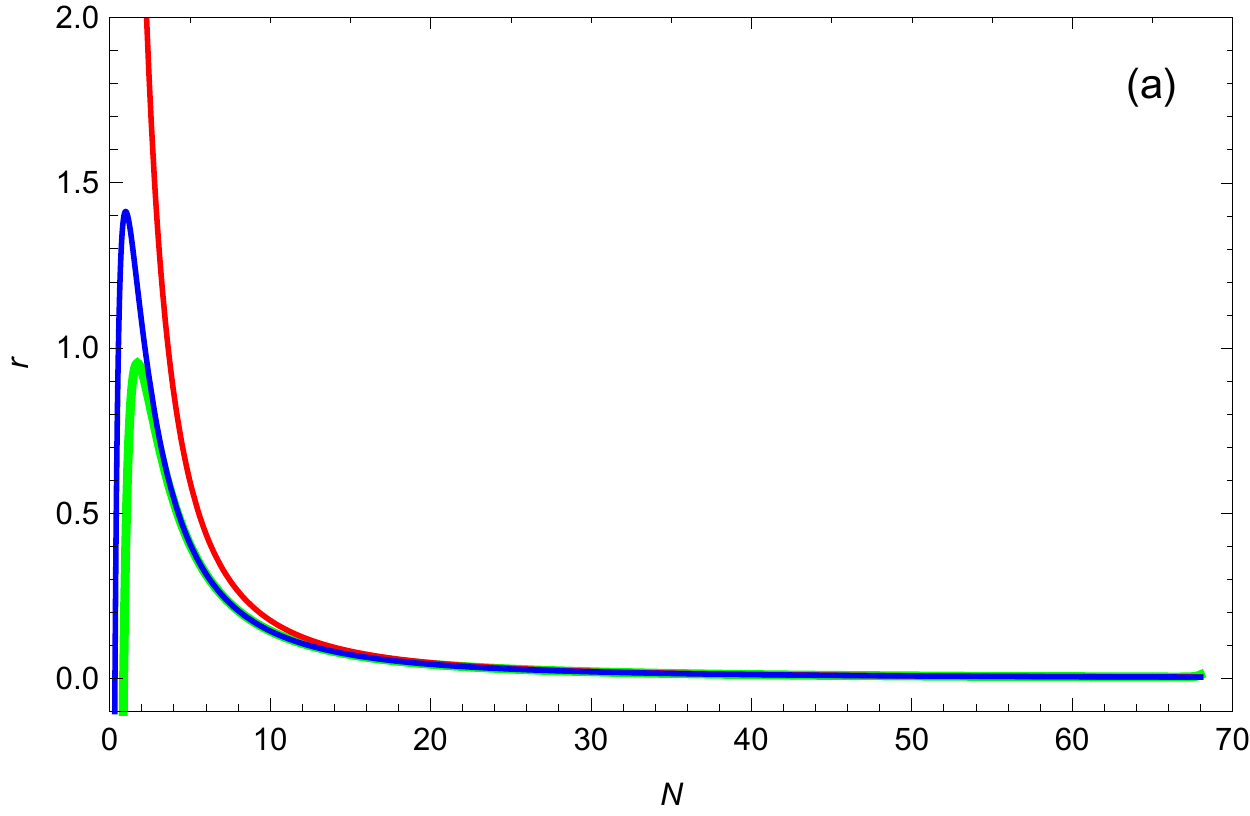}
        \hfill
        \includegraphics[width=7.65cm]{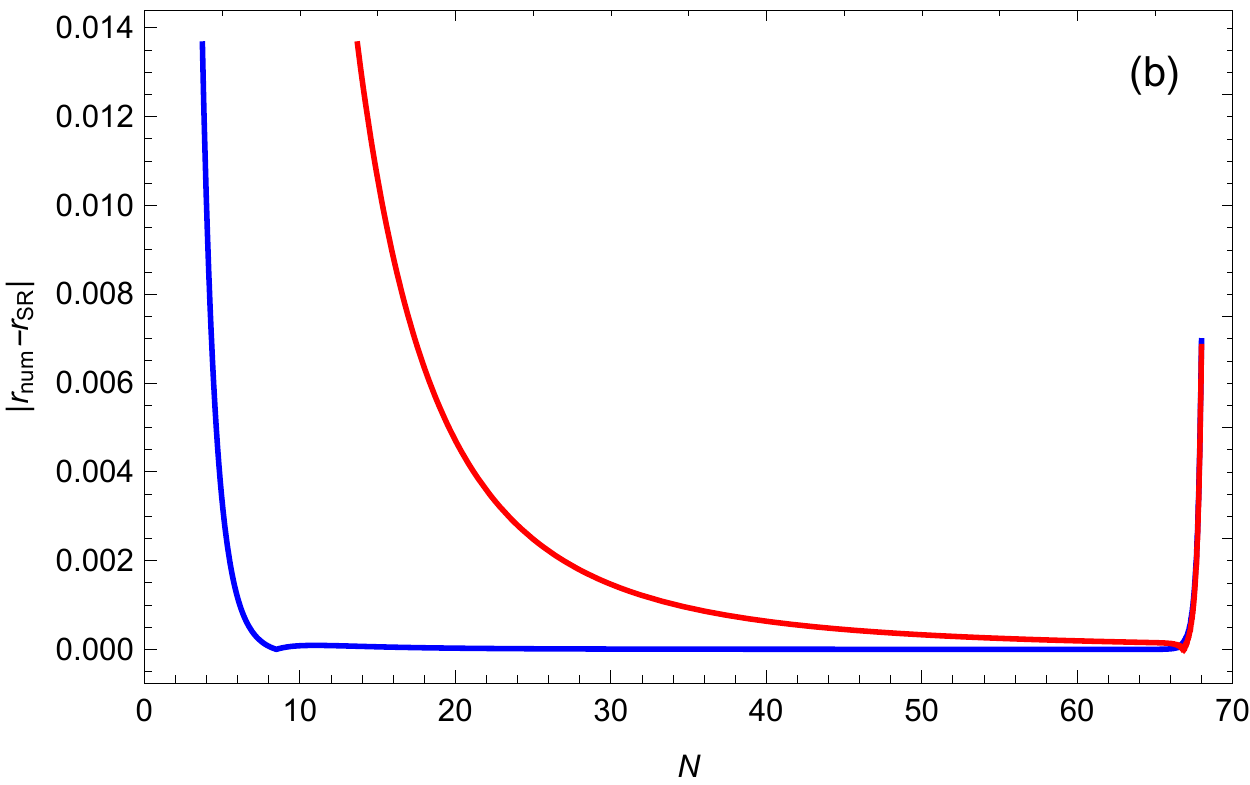}
    \caption{ (a) Tensor-to-scalar ratio $r$ for a T-model with $A = 10^{-9}$, $m=1$, $\lambda = 1/\sqrt{15}$ and initial condition
                        $h(\varphi_0)=4.4\times 10^{-5}$ for $\varphi_0=10$ as a function of the number of e-folds $N$. The red curve is the
                        result given by the first-order SR approximation, the blue curve is the result
                        given by the second-order SR approximation, and the green curve is the
                        result of a numerical integration.
                   (b) Absolute values of the differences between the numerical value of the tensor-to-scalar ratio and its
                        first order (red curve) and second order (blue curve) SR approximations. In the region beyond $N=65$ both
                         curves are superimposed.                   
	\label{fig:r}}
\end{figure}

Finally, figures~\ref{fig:ns} and~\ref{fig:r} show, respectively, the corresponding results for the spectral index $n_s$
and for the tensor-to-scalar ratio $r$ for the T-model with $m=1$ and $\lambda = 1/\sqrt{15}$. Again, the red curves
are the first-order SR results, the blue curves are the second-order SR results, and the green curves are the results of numerical
integrations. The left panels show the magnitudes, and the right panels the differences between the numerical values
and the SR approximations. In the steep arcs beyond $N=65$ in both (b) panels the first-order and second-order SR
approximations are in fact superimposed. Again, the second-order approximation is more accurate.
Both approximations quickly loose accuracy at the beginning of inflation (i.e., beyond $N=65$, well outside the SR region).
\section{Conclusions\label{sec:conc}}
We have shown that the Hamilton-Jacobi formalism, when adapted to the specific inflationary potential,
leads to efficient recurrence relations to compute asymptotic expansions
for the Hubble parameter in both the SR stage---where, in fact, the expansion corresponds to the separatrix---and in the
KD stage---where the expansion depends explicitly on the initial condition. Partial summations of these asymptotic expansions
are not accurate enough to describe the complete inflation process, but Pad\'e summations thereof
converge quickly and extend their respective domains to allow a successful matching, which in turn determines
the relation between the respective asymptotic expansions for the number of e-folds.
These SR and KD expansions combined cover the whole inflation period and are much more accurate than well known
formulas like eq.~(\ref{eq:cmrv}) for the Hubble parameter in the KD stage, or eq.~(\ref{standardSRPhi}) for the number
of e-folds during the SR stage, and allow us to find  the total amount of inflation as a function of the initial data or,
conversely, to choose initial data that correspond to a fixed total amount of inflation.
The required order of the expansions is determined by the fact that although for a fixed order the accuracy
increases with the number of e-folds, to attain a certain accuracy for a fixed number of e-folds may require
high-order expansions for the typical values of the T-models parameters $\lambda$ and $m$.
In particular, the SR expansions have allowed us to compute consistently expansions for the spectral 
index $n_s(N)$ accurate to order $1/N^2$, and the tensor-to-scalar $r(N)$ accurate to order $1/N^3$, in which we have found
logarithmic terms and the noteworthy fact that the first dependence of $n_s(N)$ on the model parameters
is found precisely in the term proportional to $1/N^2$.
\acknowledgments
The authors are grateful to Prof.~Luis Mart\'{\i}nez Alonso for useful discussions.
\appendix
\section{Adapted SR expansions\label{ap:a}}
In this Appendix we show the specific choices of $F(\Phi)$ as well as the first few terms of the expansions for $H(\Phi)$
for the families of potentials mentioned in the Introduction.
In terms of reduced variables, eqs.~(\ref{introchange}), (\ref{introymas}) and~(\ref{introiu}) read,
\begin{equation}
	\label{change}
	\mathfrak{h}(\varphi) = \frac{h(\varphi)}{\sqrt{v(\varphi)}},
\end{equation}
\begin{equation}
	\label{ymas}
	\mathfrak{h}'(\varphi) = \sqrt{\mathfrak{h}(\varphi)^2-1} - \mathfrak{v}(\varphi)  \mathfrak{h}(\varphi),
\end{equation}
\begin{equation}  
	\label{iu}
 	\mathfrak{v}(\varphi) = \frac{v'(\varphi)}{2 v(\varphi)},
\end{equation}
respectively, with
\begin{equation}
	\mathfrak{h}(\varphi) = \mathcal{H}(\Phi),
\end{equation}
\begin{equation}
	 \mathfrak{v}(\varphi) = \mathrm{M}_\mathrm{Pl} \sqrt{\frac{2}{3}} \mathcal{V}(\Phi).
\end{equation}
If $\mathfrak{v}(\varphi)$ is of the form
\begin{equation}
	\mathfrak{v}(\varphi) = Q(f(\varphi)),
\end{equation}
where $f(\varphi)\to\infty$ as $\varphi\to\infty$, and $Q$ has a suitable Taylor expansion as $f(\varphi)\to\infty$,
then eq.~(\ref{ymas}) has a formal solution in inverse powers of $f(\varphi)$.

For example, if the potential $v(\varphi)$ is a rational function of $\varphi$ (and in particular, a polynomial function), so is
$\mathfrak{v}(\varphi)$, and therefore the choice
\begin{equation}
	f(\varphi) = \varphi,
\end{equation}
is suitable for all these potentials. The same choice works for monomial potentials with non-integer exponents,
because again $\mathfrak{v}(\varphi)$ is a rational function of $\varphi$ (more concretely, proportional to $1/\varphi$).
In Table~\ref{tab:hpol} we list the first four terms of the expansions of $\mathfrak{h}(\varphi)$
for some polynomial, rational and for two monomial potentials with non-integer exponents $p$ taken from Ref.~\cite{MA14},
and in Table~\ref{tab:hpolpar} we give the relation between the parameters $A$, $a$ and $b$ in Table~\ref{tab:hpol} and the
corresponding physical parameters in Ref.~\cite{MA14}.
We have not included in the table particular instances of these polynomial potentials: For example,
MSSMI is the particular case of GMSSMI with $b=9 a^2/20$, and RIPI is the particular case of GRIPI with $b=9 a^2/32$.
\begin{table}[tbp]
\centering
\begin{tabular}{ccl}
\hline
  \mbox{Model} & $v(\varphi)$ & $\mathfrak{h}(\varphi) = \frac{h(\varphi)}{\sqrt{v(\varphi)}}$ \\
\hline
     \mbox{MLFI} &
     $A \varphi ^2 \left(a  \varphi ^2+1\right)$  &
     $1+\frac{2}{\varphi ^2}-\frac{2(a +1)}{a  \varphi^4} +\frac{24 a ^2+16 a +5}{2 a^2 \varphi ^6} - \cdots$
     \\
     \mbox{DWI}  &
     $A \left(a \varphi ^2-1\right)^2$ & $1+\frac{2}{\varphi ^2}-\frac{2 (a-2)}{a\varphi ^4} +\frac{2 \left(6 a^2-8 a+3\right)}{a^2\varphi ^6}-\cdots$
     \\
    \mbox{GMSSMI}  &
    $A \left(\varphi^2-a \varphi ^6+b \varphi ^{10}\right)$ &
    $1+\frac{25}{2\varphi ^2}+\frac{875}{8 \varphi ^4}+\frac{5 (32 a-375 b)}{16 b \varphi^6}+\cdots$
    \\
   \mbox{GRIPI} &
   $A \left(\varphi^2-a \varphi ^3+b \varphi ^4\right)$ &
   $1+\frac{2}{\varphi^2}+\frac{a}{b \varphi ^3}+
   \frac{9 a^2-16 b^2-16 b}{8 b^2 \varphi^4}+\cdots$
   \\
   \mbox{SSBI} &
   $A \left(1+a \varphi ^2+b \varphi ^4\right)$ &
   $1+\frac{2}{\varphi^2}-\frac{2(a+b)}{b \varphi ^4}+ \frac{5 a^2+16 a b+24 b^2-8 b}{2 b^2\varphi ^6}+\cdots$
   \\
   HF1I & $A (a \varphi +1)^2
   \left(1-\frac{a^2}{(a \varphi+1)^2}\right)$  & $1+\frac{1}{2\varphi ^2}-\frac{1}{a \varphi ^3}+\frac{3(a^4+4)}{8 a^2 \varphi^4}+\cdots$\\
   RGI  & $\frac{A \varphi ^2}{a+\varphi ^2}$     &     $1+\frac{a^2}{2\varphi ^6}-\frac{a^3}{\varphi ^8}
   +\frac{3 \left(a^4-2 a^3\right)}{2\varphi ^{10}} +\cdots$            \\
   CSI  & $\frac{A}{(1-a \varphi )^2}$     &   $1+\frac{1}{2\varphi ^2}+\frac{1}{a \varphi ^3}+\frac{11 a^2+12}{8 a^2 \varphi^4}+\cdots$               
        \\ 
      LFI & $A \varphi^p$  & $1+\frac{p^2}{8\varphi ^2}+\frac{p^3(3p-16)}{128\varphi ^4}
   +\frac{p^4 \left(5p^2-80p+288\right)}{1024\varphi ^{6}}+\cdots$\\
   IMI  & $A\varphi^{-p}$     &     $1+\frac{p^2}{8\varphi ^2}+\frac{p^3(3p+16)}{128\varphi ^4}
   +\frac{p^4 \left(5p^2+80p+288\right)}{1024\varphi ^{6}}+\cdots$                       \\
\hline    
\end{tabular}
\caption{\label{tab:hpol}
              First four terms of the expansions for the Hubble parameter $\mathfrak{h}(\varphi)$ as inverse power series in $f(\varphi)=\varphi$
              for polynomial, rational and monomial potentials with non-integer exponent $p$ taken from Ref.~\cite{MA14}.
              The relation between the parameters $A$, $a$ and $b$ in this table and the physical parameters in
              Ref.~\cite{MA14} is given in Table~\ref{tab:hpolpar}.}
\end{table}
\begin{table}[tbp]
\centering
\begin{tabular}{cccc}
\hline
  \mbox{Model} & $A$ & $a$ & $b$\\
\hline
    \mbox{MLFI}  &
    $\frac{2M^4}{{\mathrm{M}_\mathrm{Pl}}^2}$  &
    $\frac{2}{3}\alpha$  &
    \\
    \mbox{DWI}  &
    $\frac{3M^4}{{\mathrm{M}_\mathrm{Pl}}^2}$  &
    $\frac{2{\mathrm{M}_\mathrm{Pl}}^2}{3\phi_0^2}$   &
    \\
    \mbox{GMSSMI}  &
    $\frac{2M^4}{\phi_0^2}$  &
    $\frac{8{\mathrm{M}_\mathrm{Pl}}^4}{27\phi_0^4}\alpha$ &
    $\frac{16{\mathrm{M}_\mathrm{Pl}}^8}{405\phi_0^8}\alpha$
    \\
    \mbox{GRIPI}  &
    $\frac{2M^4}{\phi_0^2}$ &
    $\frac{4\sqrt{2}{\mathrm{M}_\mathrm{Pl}}}{3\sqrt{3}\phi_0}\alpha $ &
    $\frac{{\mathrm{M}_\mathrm{Pl}}^2}{3\phi_0^2}\alpha$
    \\
    \mbox{SSBI}  &
    $\frac{3M^4}{{\mathrm{M}_\mathrm{Pl}}^2}$ &
    $\frac{2}{3}\alpha$  &
    $\frac{4}{9}\beta$
    \\
         HF1I & $\frac{3M^4}{{\mathrm{M}_\mathrm{Pl}}^2}$  & $\sqrt{\frac{2}{3}}A_1$\\
     RGI  & $\frac{3M^4}{\mathrm{M}_\mathrm{Pl}^2}$  & $\frac{3}{2}\alpha$            \\
     CSI  & $\frac{3M^4}{\mathrm{M}_\mathrm{Pl}^2}$  & $\sqrt{\frac{2}{3}}\alpha$    \\ 
     LFI & $\left(\frac{2}{3}\right)^{\frac{p}{2}}\frac{3M^4}{\mathrm{M}_\mathrm{Pl}^2}$ \\
     IMI & $\left(\frac{2}{3}\right)^{-\frac{p}{2}}\frac{3M^4}{\mathrm{M}_\mathrm{Pl}^2}$ \\
\hline    
\end{tabular}
\caption{\label{tab:hpolpar}
              Relation between the parameters $A$, $a$ and $b$ in Table~\ref{tab:hpol}
              and the physical parameters in Ref.~\cite{MA14}.}
\end{table}

Our second set of examples pertains to potentials wherein the inflaton appears either as a rational function of an exponential
or even as the exponential of a rational function of an exponential, and comprises potentials taken from Ref.~\cite{MA14}
and $\alpha$-attractors taken form Refs.~\cite{AK18} and~\cite{CA15}. In all these cases $\mathfrak{h}(\varphi)$ can
be systematically expanded as a formal series in inverse powers of
\begin{equation}
	f(\varphi) = e^{\lambda\varphi},
\end{equation}
in some cases for a particular, fixed value of $\lambda$. Table~\ref{tab:hexp} shows the first three terms of these expansions
and Table~\ref{tab:hexppar} shows the relations between the parameters $A$ and $\lambda$ in Table~\ref{tab:hexp}
and the physical parameters in Refs.~\cite{MA14,CA15,AK18}.

An additional simplification occurs if the potential $v(\varphi)$ is an even function of $f(\varphi)$, wherein odd powers
of $1/f(\varphi)$ do not appear in the expansions and we can expand directly in $1/f(\varphi)^2$.

The key point exemplified in Tables~\ref{tab:hpol} and~\ref{tab:hexp} is that by using a suitable $f(\varphi)$ adapted to
each (family of) potential(s), a systematic rearrangement of the SR expansion as a formal series in inverse powers of
$f(\varphi)$ can be efficiently computed to any desired order, which in turn allows us the consistent use of summation
methods as discussed in the following Appendix.
\begin{sidewaystable}
\centering
\begin{tabular}{ccl}
\hline
  Model & $v(\varphi)$ & $\mathfrak{h}(\varphi)=\frac{h(\varphi)}{\sqrt{v(\varphi)}}$ \\
\hline
     HI & $A \left(1-e^{-\frac{2\varphi}{3} }\right)$  & $1+\frac{2}{9}e^{-\frac{4\varphi}{3}}+\frac{20}{81}e^{-2 \varphi
   }+\cdots$\\
    ESI & $A \left(1-e^{-\lambda  \varphi }\right)$ & $1+\frac{1}{8} \lambda ^2 e^{-2 \lambda 
   \varphi }+\frac{1}{8} \left(2\lambda ^2-\lambda ^4\right) e^{-3 \lambda\varphi }
   +\cdots$            \\
   MHI & $A\left(1-\mbox{sech}(\lambda\varphi)\right)$ & $1+\frac{\lambda ^2 }{2}e^{-2\lambda\varphi }+(2\lambda^2-\lambda^4)e^{-3 \lambda    \varphi }+\cdots$\\
   CNAI & $A(1-(1+\lambda^2)\tanh^2(\lambda\varphi))$  & $1+\frac{8 (1+ \lambda ^2)^2 }{\lambda^2}e^{-4 \lambda\varphi }
          -\frac{64\left(2 \lambda^8+6 \lambda ^6+5 \lambda ^4-1\right)
   }{\lambda^4}e^{-6 \lambda  \varphi }+\cdots$\\
   
   CNCI & $A((1+\lambda^2)\coth^2(\lambda\varphi)-1)$ & $1+\frac{8 (1+ \lambda ^2)^2 }{\lambda^2}e^{-4 \lambda\varphi }
          +\frac{64\left(2 \lambda^8+6 \lambda ^6+5 \lambda ^4-1\right)
   }{\lambda^4}e^{-6 \lambda  \varphi }+\cdots$\\
   T &    $A \left(\tanh^2(\lambda\varphi)\right)^n$  & $1+8n^2\lambda^2e^{-4\lambda\varphi}-128n^3\lambda^4e^{-6\lambda\varphi}+\cdots$\\
   E &    $A \left(1-e^{-2\lambda\varphi}\right)^{2n}$  & $1+2n^2\lambda^2e^{-4\lambda\varphi}-4n^2\lambda^2(4n\lambda^2-1)e^{-6\lambda\varphi}+\cdots$\\
  \mbox{Linear} &    $A \left(\tanh(\lambda\varphi)+1\right)+B$  & $1+\frac{2 A^2 \lambda ^2 }{(2A+B)^2}e^{-4 \lambda  \varphi }-  \frac{8A^2\lambda^2(A(2\lambda^2+1)+B)}{(2A+B)^3}e^{-6\lambda\varphi}+\cdots$\\
  \mbox{Two-shoulder} &    $A\left(\exp\left(\gamma\tanh(\lambda\varphi)\right)-1\right)^2$ & $1+\frac{8 \gamma ^2 \lambda ^2 e^{2 \gamma}}{\left(e^{\gamma}-1\right)^2}e^{ -4\lambda  \varphi }-\frac{32 \gamma ^2 \lambda ^2
   \left(e^{\gamma } \left(4 \gamma  \lambda^2+1\right)-\gamma -1\right) e^{2 \gamma  }}{\left(e^{\gamma
   }-1\right)^3}e^{ -6\lambda  \varphi }+\cdots$ \\
   \mbox{Exp-I} &    $A\exp\left(\gamma\tanh(\lambda\varphi)\right)$ & $1+2\gamma^2\lambda^2e^{-4\lambda\varphi}-8\gamma^2\lambda^2(1+2\gamma\lambda^2)e^{-6\lambda\varphi}+\cdots$            \\
   \mbox{Exp-II} &    $A\left(\exp\left(\gamma(\tanh(\lambda\varphi)+1)\right)-1\right)$ & $1+\frac{2\gamma^2\lambda^2e^{4\gamma}}{(e^{2\gamma}-1)^2}e^{-4\lambda\varphi}-\frac{8\gamma^2\lambda^2e^{4\gamma}(e^{2\gamma}(1+2\gamma\lambda^2)-\gamma-1)}{(e^{2\gamma}-1)^3}e^{-6\lambda\varphi}+\cdots$            \\
   \hline    
\end{tabular}
\caption{\label{tab:hexp}
              First three terms of the expansions for the Hubble parameter $\mathfrak{h}(\varphi)$ as inverse power series in $f(\varphi)=e^{\lambda\varphi}$
              for some potentials taken from Ref.~\cite{MA14} and for $\alpha$-attractors taken from Refs.~\cite{AK18} and~\cite{CA15}.
              In the linear model it is assumed that $B \ll A$.
              The relation between the parameters $A$ and $\lambda$ in this table and the physical parameters in
              Ref.~\cite{MA14} is given in Table~\ref{tab:hexppar}, and the relation of the parameter $B$ in the caption of Table~\ref{tab:hexppar}.}
\end{sidewaystable}

\begin{table}[tbp]
\centering
\begin{tabular}{ccc}
\hline
  Model & $A$ & $\lambda$ \\
\hline
       HI  & $\frac{3M^4}{\mathrm{M}_\mathrm{Pl}^2}$  &  \\
      ESI  & $\frac{3M^4}{\mathrm{M}_\mathrm{Pl}^2}$  & $\sqrt{\frac{2}{3}}q$            \\
      MHI  & $\frac{3M^4}{\mathrm{M}_\mathrm{Pl}^2}$  & $\sqrt{\frac{2}{3}}\frac{\mathrm{M}_\mathrm{Pl}^2}{\mu}$               \\
     CNAI and CNCI & $\frac{9M^4}{\mathrm{M}_\mathrm{Pl}^2}$  & $\frac{\alpha}{\sqrt{3}}$               \\
   T and E        & $3\alpha\mu^2$           & $\frac{1}{3\sqrt{\alpha}}$          \\
   Linear       & $3\gamma\sqrt{6\alpha}$  & $\frac{1}{3\sqrt{\alpha}}$          \\
  Exp-I & $3M^2e^{-\gamma}$       & $\frac{1}{3\sqrt{\alpha}}$           \\
   Two-shoulder and Exp-II & $3M^2e^{-2\gamma}$       & $\frac{1}{3\sqrt{\alpha}}$           \\
\hline    
\end{tabular}
\caption{\label{tab:hexppar}
             Relation between the parameters $A$ and $\lambda$ in Table~\ref{tab:hexp}
              and the physical parameters in Refs.~\cite{MA14,CA15,AK18}.
              In the Linear model, $B=3\Lambda$}
\end{table}
\section{Pad\'e approximants\label{ap:b}}
Partial sums of divergent asymptotic expansions (e.g., those obtained from
perturbation theory) usually yield very limited accuracy even when used over narrow
ranges of the independent variable. To overcome this limitation, and dating back to work in the 1970's on the
perturbation theory of the quartic anharmonic oscillator~\cite{SI70}, the use of rational approximants and in
particular of Pad\'e approximants~\cite{BGM} derived from the formal power series has proved useful in a variety of fields.
Although in some important cases it has been proved that Pad\'e approximants of increasing order ultimately converge to
the exact solution of the problem~\cite{SI70}, Pad\'e approximants are typically used on an empirical basis,
which is the approach we take here  to sum both the SR and the KD series. For completeness, we illustrate the
method in the case of the SR series~(\ref{eq:hhatfs}),
\begin{equation}
	\hat{h}_{\mathrm{SR}}(y)=\sqrt{A}\left(\frac{1-y}{1+y}\right)^m\tilde{h}_{\mathrm{SR}}(y),
\end{equation}
where we have denoted with a tilde the formal power series in $y$,
\begin{equation}
	\tilde{h}_{\mathrm{SR}}(y)=1+8m^2\lambda^2\sum_{n=2}^{\infty}(-1)^nc_ny^n.
\end{equation}
Instead of using a partial sum,
\begin{equation}
	\label{hpol}
	\tilde{h}_{\mathrm{SR}}^{[n_{\mathrm{SR}}]}(y)=1+8m^2\lambda^2\sum_{n=2}^{{n_{\mathrm{SR}}}}(-1)^nc_ny^n,
\end{equation}
which yields very limited accuracy and ultimately diverges,  we use the $[n_{\mathrm{SR}}/n_{\mathrm{SR}}+1]$
Pad\'e approximant, which is defined as the rational function with numerator of degree $n_{\mathrm{SR}}$ and
denominator of degree $n_{\mathrm{SR}}+1$ whose Taylor series as $y\to 0$ has the same $2n_{\mathrm{SR}}+1$
coefficients as the given series $\tilde{h}_{\mathrm{SR}}$, i.e., the rational function
\begin{equation}\label{hrac}\tilde{h}_{\mathrm{SR}}^{[n_{\mathrm{SR}}/n_{\mathrm{SR}}+1]}(y)=\frac{1+8m^2\lambda^2\displaystyle\sum_{j=1}^{n_{\mathrm{SR}}}(-1)^j\mu_j\,y^j}{1+8m^2\lambda^2\displaystyle\sum_{j=1}^{n_{\mathrm{SR}}+1}(-1)^j\,\nu_j\,y^j},
\end{equation}
whose coefficients $\mu_j$, $j=1,\dots,n_{\mathrm{SR}}$,  $\nu_j$, $j=1,\dots,n_{\mathrm{SR}}+1$ are such that
\begin{equation}\label{hconpade}\tilde{h}_{\mathrm{SR}}^{[n_{\mathrm{SR}}/n_{\mathrm{SR}}+1]}(y)-\tilde{h}_{\mathrm{SR}}^{[2n_{\mathrm{SR}}+1]}(y)=O(y^{2n_{\mathrm{SR}}+2})\quad\mbox{as }  y\,\rightarrow\,0.
\end{equation}
By substituting eq.~(\ref{hpol}) with $n_\mathrm{SR}$ replaced by $2n_\mathrm{SR}+1$
and eq.~(\ref{hrac}) into eq.~(\ref{hconpade}) we find,
\begin{equation}\label{padecon2}\everymath{\displaystyle}\begin{array}{l}
\left(1+8m^2\lambda^2\sum_{n=2}^{2n_{\mathrm{SR}}+1}(-1)^nc_ny^n\right)
\left(1+8m^2\lambda^2\sum_{j=1}^{n_{\mathrm{SR}}+1}(-1)^j\nu_j\,y^j\right)\\  \\
\qquad-\left(1+8m^2\lambda^2\sum_{j=1}^{n_{\mathrm{SR}}}(-1)^j\,\mu_j\,y^j\right)
=O(y^{2n_{\mathrm{SR}}+2})\quad\mbox{as }  y\,\rightarrow\,0.
\end{array}\end{equation}
Equating to zero the coefficients of $y^j$, $j=1,\dots,n_{\mathrm{SR}}$ in the left-hand side of eq~(\ref{padecon2})
we find that, with $c_1=0$,
\begin{equation}\label{mus}\mu_j=\nu_j+c_j+8m^2\lambda^2\sum_{k=1}^{j-2}c_{j-k}\nu_k,\quad j=1,\dots,n_{\mathrm{SR}},\end{equation}
while equating to zero the coefficients of $y^j$, $j=n_{\mathrm{SR}}+1,\dots,2n_{\mathrm{SR}}+1$ we find that 
the coefficients $\nu_j$, $j=1,\dots,n_{\mathrm{SR}}+1$ are the solutions of the linear system,
\begin{equation}\everymath{\displaystyle}\label{nus}
\left[\begin{array}{ccccc}
8m^2\lambda^2c_{n_{\mathrm{SR}}}  & 8m^2\lambda^2c_{n_{\mathrm{SR}}-1}& \cdots &8m^2\lambda^2c_1 & 1 \\
8m^2\lambda^2c_{n_{\mathrm{SR}}+1}& 8m^2\lambda^2c_{n_{\mathrm{SR}}}  & \cdots &8m^2\lambda^2c_2 & 8m^2\lambda^2c_1\\
\vdots     & \vdots     &        &\ddots& \vdots \\
8m^2\lambda^2c_{2n_{\mathrm{SR}}} &8m^2\lambda^2c_{2n_{\mathrm{SR}}-1}& \cdots &8m^2\lambda^2c_{n_{\mathrm{SR}}+1}&
8m^2\lambda^2c_{n_{\mathrm{SR}}}
\end{array}\right]
\left[\begin{array}{c} \nu_1 \\ {\nu_2} \\ {\vdots} \\ \nu_{n_{\mathrm{SR}}+1}\end{array}\right]\,=\,
\left[\begin{array}{c} -c_{n_{\mathrm{SR}}+1} \\ -c_{n_{\mathrm{SR}}+2} \\ {\vdots} \\ -c_{2n_{\mathrm{SR}}+1}\end{array}\right].
\end{equation}

For example, for $n_{\mathrm{SR}}=1$ we find that $\mu_1=\nu_1=-\frac{2}{m}$, $\nu_2=-1$ and therefore
\begin{equation}
	\tilde{h}_{\mathrm{SR}}^{[1/2]}(y)=\frac{1+16m\lambda^2y}{1+16m\lambda^2y-8m^2\lambda^2y^2}.
\end{equation}

The choice of $n_{\mathrm{SR}}$ and $n_{\mathrm{SR}}+1$ as degrees of the polynomials in the numerator and
in the denominator of the approximant~(\ref{hrac}) is not critical (in most cases, any para-diagonal sequence of approximants
can be used), and has been chosen for efficiency in the recursive solution of the system~(\ref{nus}).
The accuracy of this approximants as compared with numerical solutions of the corresponding equations is
discussed in the main body of the paper.
\bibliographystyle{JHEP}
\bibliography{tmodels}
\end{document}